\documentclass[preprintnumbers,superscriptaddressz,amsmath,floatfix,amssymb,prd]{revtex4}
	
	\usepackage{amsmath,amssymb,bm}
	\usepackage{graphicx}
	\usepackage{hyperref}
	\usepackage{amsfonts}
	\usepackage{latexsym}
	\usepackage{amsmath}
	\usepackage{amssymb}
	\usepackage{slashed}
	\usepackage{graphicx}
	\usepackage{caption}
	\usepackage{subcaption}
	\usepackage{tikz}
	\usetikzlibrary{arrows,decorations.pathmorphing,backgrounds,positioning,fit,petri}
	\usetikzlibrary{arrows.meta}
	\usepackage{yfonts}
	\usepackage{color}
	\newcommand{\beq}{\begin{eqnarray}}
	\newcommand{\eeq}{\end{eqnarray}}

\begin{document}

		\title{Hydrodynamic Excitations from Chiral Kinetic Theory and the Hydrodynamic Frames}
		
		\author{Navid Abbasi}
		
		\affiliation{School of Particles and Accelerators, Institute for Research in Fundamental Sciences (IPM), P.O. Box 19395-5531, Tehran, Iran}


		\author{ Farid Taghinavaz}
		
		\affiliation{School of Particles and Accelerators, Institute for Research in Fundamental Sciences (IPM), P.O. Box 19395-5531, Tehran, Iran}
		
			\author{Kiarash Naderi}
		\affiliation{Sharif University of Technology, Tehran, Iran}

		\begin{abstract}
			{In the framework of chiral kinetic theory (CKT), we consider a system of right- and left-handed Weyl fermions out of thermal equilibrium in a homogeneous weak magnetic field.   We show that the Lorentz invariance implies a modification in the definition of the momentum current in the phase space, compared to the case in which the system is in global equilibrium.  Using this modified momentum current, we derive   the linearized  conservation equations from the kinetic equation  up to second order in the derivative expansion.  It turns out that the eigenmodes of these equations, namely the hydrodynamic modes, differ from those obtained from the hydrodynamic in the Landau-Lifshitz (LL) frame at the same order. We show that the modes of the former case may be transformed to the corresponding modes in the latter case by a global boost. The velocity of the boost is proportional to the magnetic field as well as the difference between the right- and left-handed charges susceptibility.  We then compute the chiral transport coefficients in a system of non-Abelian chiral fermions in the no-drag frame and by making the above boost,  obtain the well-known transport coeffiecients of the system in the LL frame.  Finally by using the idea of boost, we reproduce the  AdS/CFT result for the chiral drag force exerted on a quark at rest in the rest frame of the fluid, without performing any holographic computations. }
			
		\end{abstract}
		\maketitle

	\section{Introduction}
	\label{1}
The parity-odd transport in chiral systems has attracted much interest recently. The generation of non-dissipative currents in the direction  of external magnetic field (chiral magnetic effect (CME)) can be related to presence of microscopic anomaly in such systems \cite{Vilenkin:1980fu,Ninomia}.   When this effect is accompanied with the chiral separation effect, would lead to propagation of a gap-less excitation in the system, the so-called chiral magnetic wave (CMW) \cite{Kharzeev:2010gd}. In heavy ion collisions, CMW  induces the electric quadrupole moment of the quark-gluon plasma \cite{Burnier:2011bf,Adamczyk:2015eqo,Belmont:2014lta}. In condensed matter the CME may explain the negative magneto-resistance in Weyl semimetals \cite{Weylsemimeta:Son,Weylsemimeta} and the same in Dirac semimetals \cite{Diracsemimeta}. On the other hand, the anomalous current may be induced in the direction of rotation as well. The well-known chiral vortical effect (CVE) was first discussed in the context of astrophysics  \cite{Vilenkin:1979ui}. 

The chiral transport has been studied in various theoretical frameworks.
 In lattice field theory, CMW has been numerically shown to exist in gauge theory 	\cite{Buividovich:2009wi}.   In AdS/CFT, CVE  has been found in the long wavelength limit of boundary gauge theory, i.e. in the hydrodynamics regime \cite{Erdmenger:2008rm,Banerjee:2008th}. The latter, specifically,  motivated  to construct the chiral hydrodynamics \cite{Son:2009tf,Jensen:2012jh,Banerjee:2012iz}.  
Recently, the presence of mixed gauge-gravitational anomaly  predicted by chiral hydrodynamics, has been  confirmed	in Weyl semimetals 	\cite{Gooth:2017mbd}. 

Kinetic theory is another theoretical framework by which one can study the non-equilibrium dynamics in classical systems with rare collisions. Interestingly, it can be used to study the chiral transport in a system of Weyl fermions in presence of electromagnetic fields.
If the flux of Berry curvature through a given Fermi surface is non-vanishing, the particle number associated with the Fermi surface has a triangle anomaly 	\cite{Son:2012wh}.  So one can enter the effect  of anomalous particle number non-conservation in kinetic equation by putting a  Berry monopole at the origin of the momentum space. The resultant kinetic description is called the chiral kinetic theory 	\cite{Stephanov:2012ki}.
Using chiral kinetic theory, one can find the non-equilibrium expressions for the CME as well as the CVE in a system of Weyl fermions 	\cite{Stephanov:2012ki}.

Collective excitations are basically the macroscopic observables reflecting the microscopic dynamics in the system, specifically the effect of anomalies,  at low energy.
To get more insight about the chiral transport from chiral kinetic theory, it is natural to study its collective excitations, specially in the long wavelength limit, corresponding to the hydrodynamics.  Such study for a system of right- and left-handed fermions, coupled to an external homogeneous magnetic field, was first done in \cite{Stephanov:2014dma}. In the mentioned paper, the collective modes either in long wave-length regime and also beyond the hydro regime were found. However, their study was under this approximation that the energy and momentum perturbations are decoupled from the chiral charge perturbations. Considering only the perturbations of two opposite chirality charge densities, they derived the expression of chiral magnetic wave from the chiral kinetic theory. 
As the first part in this paper we want to extend the computations of  \cite{Stephanov:2014dma} to the case in which all the hydrodynamical perturbations are taken into account. While at first look it seems straightforward, we will show that such extension requires a new modification in the structure of kinetic theory. 

Let us recall that the classical action of the massless Dirac particles is not manifestly Lorentz invariant \cite{Chen:2014cla}. It has been shown that Lorentz invariance dictates  a modification in the energy dispersion of massless particles in magnetic field \cite{Son:2012zy}. It is exactly this modification which ensures the angular momentum conservation in the collisions and also leads to non-locality in the collision integral due to the side jump 	\cite{Chen:2015gta}. The above-mentioned shift in the dispersion of energy is specific to global equilibrium in which the system can either couple to an external  constant magnetic field or uniformly rotate around an axis. 

When considering  the system out of equilibrium, however, local vorticity  induced by the momentum perturbations may play a non-trivial role in the system. We will extend the computations given in \cite{Son:2012zy} to the latter case and show that the Lorentz invariance obliges to modify the momentum current of the massless particles, out of equilibrium. While the shift in the energy dispersion of the particles found in \cite{Son:2012zy} is due to a spin-magnetic coupling, in our case, the shift in momentum current turns out to be due to spin-vorticity coupling. Using this modified momentum current, we then find the set of conservation equations from the kinetic equation, in the long wave length regime and in a weak magnetic field.  In this setup, one naturally expects to observe the mixing between different perturbations. We show that CMWs will mix with the perturbations of  temperatue and make the chiral magnetic heat waves. The sound wave will mix with perturbations of charge densities as well. Moreover, the magnetic field may couple with the local perturbations of vorticity in the system and give chiral Alfv\'en waves, (see \cite{Yamamoto:2015ria,Abbasi:2015saa,Abbasi:2016rds} for first observations of chiral Alfv\'en waves in the hydrodynamics).

Independently and in the context of chiral hydrodynamics, a similar study in a chiral system had been done before. In  \cite{Abbasi:2016rds},  hydrodynamic excitations in a rotating chiral fluid coupled to a magnetic field was computed. While all the computations in  \cite{Abbasi:2016rds} are done for a general conformal fluid, the final results are reported for a system of weakly coupled chiral fermions. Such system is exactly the one under our study in the present paper. So we would consider the case without global vorticity in  \cite{Abbasi:2016rds} and compare their results with what we obtain from chiral kinetic theory. It is seen that the expressions of six hydrodynamic modes in the above two cases are not exactly the same. Each of the dispersion relations obtained in \cite{Abbasi:2016rds} differs from its counterpart in chiral kinetic theory, by a contribution proportional to the magnetic field.

In the second part of the paper we will focus on realizing the reason for the above difference. Since in both cases mentioned above the magnetic field is taken so weak as a one derivative object, one notices that the difference between two sets of results is just due to the derivative corrections. It is basically reminiscent of the  difference between the different fluid frames in hydrodynamics. As it is well-known, out of local equilibrium, namely beyond the zero order in hydrodynamic expansion, the hydrodynamic fields are not well-defined. One can redefine velocity, temperature, etc, by freely adding any symmetry-allowed combination of derivative terms. By making each redefinition, however, the physical objects like energy density and charge density remain invariant. Each choice  of defining the hydro fields is also referred to as a frame in hydrodynamics. 

Our main idea to explain the discrepancy between the two sets of results is that their associated computations are being done in two different hydrodynamic frames.  The frame used in \cite{Abbasi:2016rds} is the Landau-Lifshitz (LL)  frame. By definition, it is the  hydrodynamical frame in which no energy flows in the rest frame of the fluid, namely  $u^{\mu}=(1,\boldsymbol{0})$ means no energy current. In any another frame, the energy flow in the rest frame of the fluid may then be only proportional to the magnetic field (or global vorticity), when neglecting the dissipation.   By explicit calculations in kinetic theory side we show that the latter is the case for our system of chiral fermions \footnote{Since we are
working in the non-dissipative regime, no other one derivative data, like expansion, could contribute to the energy flow in rest frame.}; when $u^{\mu}=(1,\boldsymbol{0})$, an energy flow proportional to the magnetic field exists in the system. 
This confirms that the two sets of computations explained in previous paragraph correspond to two differnt hydrodynamic frames.

Interestingly, since the magnetic field is a vector itself, we expect by changing the frame from the LL to the chiral kinetic theory only the velocity vector to be corrected.  On the other hand, the magnetic field is non-vanishing in the equilibrium in both frames.  We conclude that in the equilibrium  state,  the rest frame of the fluid in the LL frame may be transformed to that in  our system by a boost proportional to the magnetic field.
We  compute the velocity of the associated boost for a system which is coupled to a magnetic field and is unifromly rotating around an axis.

   The idea of boost in equilibrium state is in fact limited to cases that at least one non-vanishing  one derivative object, like magnetic field, exists in equilibrium.  
In the absence of such objects one can still redefine the hydro fields by one derivative dissipative data, however, the different frames would not be distinguished from each other in equilibrium.  This is exactly the case in deriving the Eckart frame from the  LL
  frame \cite{Kovtun:2012rj}.  Since the energy flow  vanishes in the rest frame of fluid in both Eckart and LL frames,   these two frame are not distinguishable in the equilibrium.
  
Let us recall that the collective excitations propagate on top of the equilibrium state in the rest frame of the fluid.
Knowing this and based on the above arguments, one can understand how the hydrodynamic modes obtained in this paper would be related to the modes in the LL frame.  We show that the velocity of each modes in former frame is transformed to velocity of its counterpart in latter frame by exactly the above-mentioned boost between the rest frames in equilibrium.        

On the other hand, the hydrodynamic regime of chiral kinetic theory corresponds to the hydrodynamic in the Laboratory (Lab) frame \cite{Landsteiner:2016led} (sometimes it is also called the thermodynamic frame \cite{Jensen:2012jh}). This fluid frame is the one in which the rest frame of the fluid in equilibrium coincides with the laboratory Lorentz frame to all orders  in derivative expansion. 
In a chiral system with only axial current, the Lab frame has been shown to have this interesting property that a heavy impurity at rest in the rest frame of the fluid is not dragged by the chiral fluid \cite{Stephanov:2015roa}. For this reason, the Lab frame is called a no-drag frame as well.

Using the idea of boost,  one can find the components of the energy-momentum tensor and charge currents in the rest frame of the Lab frame from those of the LL frame and vice versa.
On the other hand, using the second law of thermodynamics, the constitutive relations for a system of non-Abelian chiral fermions have been computed in the LL frame in \cite{Neiman:2010zi}.
So as the second check of the idea of boost, we first compute the anomalous transport coefficients in the no-drag frame for a system of non-Abelian chiral fermions. Then by applying the boost, we reproduce the results of  \cite{Neiman:2010zi}. Let us denote that one can reduce the non-Abelian group to $U(1)_V\times U(1)_A$. The latter may correspond to a system of right- and left-handed chiral fermions, the one we discussed earlier in kinetic theory. We show that the transport coefficients of the latter system \cite{Landsteiner:2012kd,Landsteiner:2016led,Gao:2012ix}, are a special subclass of  the no-drag frame results. This could be regarded as the simple generalization of the \cite{Stephanov:2015roa} to a chiral system with two types of chirality.
 As expected, the hydrodynamic limit of chiral kinetic theory  corresponds to a no-drag frame. 
	
	At the last part of the paper we give another evidence in favor of the boost idea. As it is well-known, the  drag force exerted on a heavy quark in a  fluid can be computed in the context of AdS/CFT. Based on the seminal computations of \cite{Gubser:2006bz,Herzog:2006gh} and by the use of Fluid/Gravity duality \cite{Bhattacharyya:2008jc}, the authors of \cite{Rajagopal:2015roa}
	computed the drag force exerted on a moving quark in a chiral fluid in presence of a constant magnetic field  in the LL frame. We will show that a part of the computations of   \cite{Rajagopal:2015roa} associated with the chiral drag force affected on a quark at rest in the rest frame of the fluid can be independently found just by the above-mentioned  boost together with some previously-reported results of Fluid/Gravity duality.  To show that, we first find the most general form of the chiral drag force in a fluid up to first order in derivative expansion. We rewrite this general expression for a moving quark in the rest frame of the fluid in LL frame and then make a boost to go to  the rest frame of the fluid in the  no-drag frame.
	Requiring the resultant drag force to vanish for a quark at rest in the new frame, we find two constraints on the coefficients of general drag force mentioned above. We show that these two constrains together with the coefficient of drag force at zero order \cite{Gubser:2006bz,Herzog:2006gh} and also with the expressions of two chiral energy transport coefficients \cite{Landsteiner:2012kd} are sufficient to determine the chiral drag force computed in  \cite{Rajagopal:2015roa} for a quark at rest in the rest frame of the fluid in LL frame. Let us emphasize that we reproduce this result without performing any AdS/CFT computations.

	The paper is organized as it follows; in section \ref{2}, after a brief review of chiral kinetic theory we show how the Lorentz invariance implies a modification in the definition of the momentum current. In section \ref{3} we compute the spectrum of hydrodynamic modes from chiral kinetic theory. In section \ref{4} we review the hydrodynamic of chiral systems and review some of the results of \cite{Abbasi:2016rds}. In section \ref{sec_5}, we motivate the idea of boost and make relation between the results of section \ref{3} with those reviewed in section \ref{4}. In section \ref{sec_6} we will show that the hydrodynamic frame corresponding to the chiral kinetic theory is a no-drag frame. In section \ref{7}, we utilize the idea of boost from section \ref{sec_5} and reproduce the chiral drag force computed via AdS/CFT for a quark at rest in the rest frame of the fluid in the LL frame. We end in section  with reviewing the novel results and pointing some future directions.

	\section{ Chiral Kinetic Theory and its new Modification}
	\label{2}
	In a system of weakly interacting quasi-particles with rare collisions, the out of equilibrium dynamics can be described by the evolution of the distribution function of quasi-particles, $n_{\textbf{p}}$,  in $(\textbf{x},\textbf{p})$ space. The dynamics of distribution function is governed by the 
the Boltzmann transport equation
	\begin{equation}\label{kinetic_eq}
	\frac{\partial n_\textbf{p}}{\partial t}+\dot{\textbf{x}}\cdot\frac{\partial n_{\textbf{p}}}{\partial \textbf{x}}+\dot{\textbf{p}}\cdot\frac{\partial n_\textbf{p}}{\partial\textbf{p}}=I_{coll}\{n_{\textbf{p}}\},
	\end{equation}
where $I_{coll}\{n_{\textbf{p}}\}$ is the collision integral.
In order to find $n_{\textbf{p}}$ from the kinetic equation, we have to add the equations of motion for $\textbf{x}$  and $\textbf{p}$  to the above equation. In presence of a homogeneous electromagnetic field,  we may write the equations for the Weyl fermions as
	\begin{eqnarray}\label{eq_motion_x}
	\dot{\textbf{x}}&=&\frac{\partial \epsilon_{\textbf{p}}}{\partial \textbf{p}}+ \dot{\textbf{p}}\times \boldsymbol{\Omega}_{\textbf{p}},\\\label{eq_motion_p}
	\dot{\textbf{p}}&=&  e \, \textbf{E}+ e\,\dot{\textbf{x}}\times \textbf{B}.
	\end{eqnarray}
	Here $\boldsymbol{\Omega}_{\textbf{p}}$ is the Berry flux which is the flux corresponding to the Berry monopole located at the center of momentum space  and is  given by
	\begin{equation}\label{Berry}
	\Omega_{\textbf{p}}=\boldsymbol{\nabla}_{\textbf{p}}\times \textbf{A}_{\textbf{p}}=\lambda\,e\,  \frac{\textbf{p}}{\text{p}^3},\,\,\,\,\,\,\,\,\,\,\,\,\,\,\,\,\textbf{A}_{\textbf{p}}=i\langle u_{\textbf{p}}|\nabla_{\textbf
		p}u_{\textbf{p}}\rangle.
	\end{equation}
	In the above expression,  $\lambda=\pm1/2$ is the helicity (or equivalently the chiralitry) of right- and left-handed Weyl fermions. 
	Right particles with $\lambda=1/2$ are sometimes referred to as $\chi=R$ and left particles $\lambda=-1/2$ as  $\chi=L$ in this paper. 
	The presence of $e=\pm1$ in the expression of  \eqref{Berry} shows that the Berry curvature for particles and anti particles have opposite signs.
	Combining the equations (\ref{eq_motion_x} and  \ref{eq_motion_p}), we may write
	\begin{eqnarray}\label{equation-Berry}
	\sqrt{G}\dot{\textbf{x}}&=&\frac{\partial \epsilon_{\textbf{p}}}{\partial \textbf{p}}+e\, \textbf{E}\times \boldsymbol{\Omega}_{\textbf{p}} + e\, \textbf{B} \left(\frac{\partial \epsilon_{\textbf{p}}}{\partial \textbf{p}}\cdot\,  \boldsymbol{\Omega}_{\textbf{p}} \right),\nonumber\\
	\sqrt{G}\dot{\textbf{p}}&=& e\, \textbf{E}+ e \, \frac{\partial \epsilon_{\textbf{p}}}{\partial \textbf{p}} \times \textbf{B} + e^{2}\,\,  \boldsymbol{\Omega}_{\textbf{p}} \left(\textbf{E}\cdot\, \textbf{B}\right),
	\end{eqnarray}
	where $G=(1+\, e\textbf{B}\cdot \boldsymbol{\Omega}_{\textbf{p}})^2$ is the phase space invariant.	While for a massless right-handed Weyl particle we  simply expect the energy disperses as $\epsilon(\textbf{p})=\text{p}$,   the presence of an external magnetic field changes this picture. It has been shown that if the Weyl particles couple to magnetic field,  Lorentz invariance requires a shift of energy in the dispersion relation equal to the spin magnetic moment coupling \cite{Son:2012zy}. It can be simply computed as the following. First we define energy density as 
\begin{equation}
T^{00}\equiv\epsilon=\int\frac{d^3\text{p}}{(2\pi)^3}\sqrt{G}\,\,\epsilon(\textbf{p})\,n_{\mathbf{p}}.
\end{equation}
	Then we multiply equation \eqref{kinetic_eq} by $\sqrt{G} \epsilon(\textbf{p})$
 and then integrate over $\textbf{p}$ to write it as $\partial _{\mu}T^{\mu 0}=E^i j^i$ with
 \begin{equation}\label{T0i}
 T^{i0}=-\int \frac{d^3 \text{p}}{(2\pi)^3}\left[(\delta^{ij}+e\,\text{B}^i\Omega^j)\frac{\epsilon_{\textbf{p}}^2}{2}\frac{\partial n_{\textbf{p}}}{\partial \text{p}^j}+\epsilon^{ijk}\frac{\epsilon_{\textbf{p}}^2}{2}\Omega^j\frac{\partial n_{\textbf{p}}}{\partial \text{x}^k}\right].
 \end{equation}
  Lorentz invariance obliges that $T^{i0}$ has to be equal to the momentum density defined as
 \begin{equation}\label{Ti0}
 T^{0i}\equiv \pi^i=\int\frac{d^3\text{p}}{(2\pi)^3}\sqrt{G}\,\,\tilde{\text{p}}_i\,n_{\mathbf{p}}.
 \end{equation}
 Here $\tilde{\text{p}}$ as the modified momentum in phase space. In a homogeneous system, $\frac{\partial n_{\textbf{p}}}{\partial \textbf{x}}=0$, Lorentz invariance up to first order in weak magnetic field gives \cite{Son:2012zy}
 \begin{equation}\label{epsilon_son}
	\epsilon(\textbf{p})=\text{p}- \lambda\,e\,\frac{\textbf{B}\cdot\textbf{p}}{\text{p}^2},\,\,\,\,\,\,\,\,\,\tilde{\textbf{p}}=\textbf{p}.
	\end{equation} 
In the following we will see that as far as we study the Weyl systems in global equilibrium these modifications are sufficient. However, when the system is deviated from its global equilibrium, the momentum will be shifted as well.
	
	\subsection{Global Equilibrium State }
	
		Since the theory of Weyl fermions is invariant under CPT, the antiparticles participate in collisions as well and consequently, we must take them into account.
	We can do it by considering the particle charge $e=\pm1$ as an additional discrete index of the distribution function. 
	 According the above considerations, the equilibrium distribution function
	of fermions is given by: 
	\begin{equation}\label{equi_dis_func}
	\tilde{n}_{\textbf{p}}^{(\lambda,e)}=\frac{1}{e^{\beta (\epsilon(\textbf{p})-e\mu_{\chi})}+1}.
	\end{equation}
	Since the energy dispersion of the particles given by \eqref{epsilon_son} depends on helicity, we consider $\lambda$ as another index of the distribution function. In the above relation, $\chi=R,L$ refers to the right- and left-handed particles (and anti-particles) respectively.
	
	In this work our system will be always subject to an external weak magnetic field. The strength of the magnetic field is of the order of derivatives of quantities, near  equilibrium. Therefore one can expand the above distribution function in a derivative expansion as the following
	\begin{equation}\label{dis_func_thermo}
	\tilde{n}^{(\lambda,e)}_{\textbf{p}}=		\tilde{n}^{(\lambda,e)}_{0\textbf{p}}-\left(\frac{\partial 	\tilde{n}^{(\lambda,e)}_{0\textbf{p}}}{\partial \epsilon(\textbf{p})}\right)_{eq.}\left( \lambda\,e\,\frac{\textbf{B}\cdot\textbf{p}}{\text{p}^2}\right) \epsilon_{f}+O(\epsilon_f^2),
	\end{equation}
	with $\tilde{n}^{(\lambda,e)}_{0\textbf{p}}=\left(e^{\beta(\text{p}-e \mu_\chi)}+1\right)^{-1}$. Please note that $\left(\frac{\partial 	\tilde{n}^{(\lambda,e)}_{0\textbf{p}}}{\partial \epsilon(\textbf{p})}\right)_{eq.}$ is taken with respect to the $\epsilon(\textbf{p})= \text{p}$.  Here $\epsilon_f$ is a parameter which counts the number of derivatives and we set it to unity at the end of the computations. It is clear that the equation \eqref{dis_func_thermo} gives us the most general thermodynamic distribution function of a system of single right- and left- handed fermions in presence of long wavelength background fields, perturbatively,  up to first order in derivative expansion. 
	
	Using the above distribution function, one may compute the following thermodynamic quantities:
	\begin{eqnarray}\label{thermo}
	w=\,\frac{4}{3}\,(\epsilon_{R}+ \epsilon_{L}) =\,\frac{4}{3}\sum_{\lambda} \sum_{e}\int_{\textbf{p}}\,\epsilon(\textbf{p})\,\tilde{n}_{\textbf{p}}^{(\lambda,\, e)}&=&\,T^4 \left(\frac{\mu_{R}^4+ \mu_{L}^{4}}{6 \pi ^2 T^4}+\frac{\mu_{R}^2+ \mu_{L}^{2}}{3 T^2}+\frac{7 \pi ^2}{45}\right),\nonumber\\
	n_{\chi}=\,\,\,\sum_{e}\int_{\textbf{p}} \,\, e\, \tilde{n}_{\textbf{p}}^{(\lambda,\, e)}\,\,&=&\,T^3 \left(\frac{\mu_{\chi}^3}{6 \pi ^2 T^3}+\frac{\mu_{\chi} }{6 T}\right),\,\,\,\,\,\,\,\,\,\,\chi=R,L\nonumber\\
	\bar{\boldsymbol{\chi}}_{\chi}\,=\, \left(\frac{\partial \,n_{\chi}}{\partial \mu_{\chi}}\right)_{T}&=& \,T^2 \left(\frac{\mu_{\chi}^2}{2 \pi ^2 T^2}+\frac{1}{6}\right),\,\,\,\,\,\,\,\,\,\,\,\,\,\,\,\,\chi=R,L\nonumber\\
	C_{v}= \frac{\partial \epsilon}{\partial T} &=& \frac{7 \pi^{2} T^{3}}{15} + \frac{T \left(\mu_{R}^{2} + \mu_{L}^{2}\right)}{2},
	\end{eqnarray}
		where we have used the notation
	$	\int_{\textbf{p}}\equiv\int \frac{d^3 \text{p}}{(2\pi)^3}\sqrt{G}$.	 Please note that while $\chi$ denotes the chirality in this paper, $\bar{\boldsymbol{\chi}}_{\chi}$ refers to the thermodynamic susceptibility associated with the charge $n_{\chi}$.
	\subsection{Local Equilibrium State}
	Out of global equilibrium, the thermodynamic variables like $\beta$, $\mu$ and $\boldsymbol{u}$ (the velocity of the system which vanished in equilibrium in Lab frame) will no longer be constant in the whole of the system and they are not even meaningful quantities. In the long wavelength limit however, in the absence of dissipative processes,  one would be able to  promote them to be local functions of space and time and write the distribution function in the local co-moving frame as
	\begin{equation}\label{local_distru}
	n^{(\lambda,e)}_{\textbf{p}}(\textbf{x},t)=\frac{1}{e^{\beta(\textbf{x},t)\big(\epsilon(\textbf{p})-\textbf{p}\cdot\boldsymbol{u}(\textbf{x},t)-\lambda\,e\, \hat{\textbf{p}}\cdot\boldsymbol{\omega}(\textbf{x},t)-e\mu_{\chi}(\textbf{x},t)\big)}+1}.
	\end{equation}
	If the length scale of variations in the system is much larger than the microscopic length scale in the system,
	$\beta(\textbf{x},t)^{-1}$, $\boldsymbol{u}(\textbf{x},t)$, $\boldsymbol{\omega}(\textbf{x},t)=\frac{1}{2} \boldsymbol{\nabla}\times \boldsymbol{u}(\textbf{x},t)$ and  $\mu_{\chi}(\textbf{x},t)$ identify with the  temperature,  velocity,  vorticity and  chemical potential,  all as the local functions of space and time, respectively.
	These functions are called the hydrodynamic fields in local equilibrium state.
	
	One may have noticed that \eqref{local_distru} solves the kinetic equation at zero order in derivative expansion. Thanks to derivative expansion in long wavelength limit, we can also find the solution at first order by expanding the hydrodynamic fields around their thermodynamic values. To proceed, we may write
	\begin{equation}\label{fluc}
	\beta(\textbf{x},t)=\beta+\epsilon_F\; \delta\beta(\textbf{x},t),\,\,\,\,\,\mu_{\chi}(\textbf{x},t)=\mu_{\chi}+\epsilon_F \;\delta \mu_{\chi}(\textbf{x},t),\,\,\,\,\,\boldsymbol{u}(\textbf{x},t)=\epsilon_F \;\frac{\boldsymbol{\pi}(\textbf{x},t)}{w}
	\end{equation} 
	with $\epsilon_F$ denoting another perturbative parameter which counts the order of fluctuations around equilibrium and will be set to unity at the end of computations. Due to later requirements we have defined the fluid velocity fluctuations in terms of momentum fluctuations $\boldsymbol{\pi}$ and the enthalpy density $w$.  From now on we drop the arguments of the fluctuation fields and work with $\delta \beta$, $\boldsymbol{\pi}$ and $\delta \mu_{\chi}$ for simplicity.
	
	In order to specify the solution of the kinetic equation at first order, we now use  \eqref{fluc}  and    expand \eqref{local_distru} around the equilibrium solution \eqref{equi_dis_func}. To first order in fluctuations, we obtain:
	\begin{equation}\label{expanded distribu}
	\begin{split}
	n^{(\lambda,e)}_{\textbf{p}}(\textbf{x},t)=&\,\tilde{n}_{\textbf{p}}^{(\lambda,e)}+\left(\frac{\partial 	\tilde{n}_{\textbf{p}}^{(\lambda,e)}}{\partial \epsilon(\textbf{p})}\right)_{eq.}\left( \frac{\epsilon(\textbf{p})-e\mu_\chi}{\beta}\delta \beta-\left(\textbf{p} + \frac{\lambda\,e}{2} \hat{p} \times \boldsymbol{\nabla}\right)\cdot \frac{\boldsymbol{\pi}}{w}-e\delta \mu_\chi \right)\epsilon_{F}+O(\epsilon_F^2)\\
	=&\,\tilde{n}^{(\lambda,e)}_{0\textbf{p}}-\left(\frac{\partial 	\tilde{n}^{(\lambda,e)}_{0\textbf{p}}}{\partial \epsilon(\textbf{p})}\right)_{eq.}\left(-e\,\lambda\,\frac{\textbf{B}\cdot\textbf{p}}{\text{p}^2}+ \frac{\epsilon(\textbf{p})-e\mu_\chi}{\beta}\delta \beta-\left(\textbf{p} + \frac{\lambda\,e}{2} \hat{p} \times \boldsymbol{\nabla}\right)\cdot \frac{\boldsymbol{\pi}}{w}-e\delta \mu_\chi\right)+O(\epsilon^2_{f},\epsilon_F^2),
	\end{split}
	\end{equation}
	where in writing the second line we have used \eqref{dis_func_thermo}. Let us note that among the corrections in the above expression, the first term in parentheses is just a derivative term of order $O(\epsilon_f \epsilon_F^0)$. While the second and fourth terms are of order $O(\epsilon_f^0\epsilon_F)$, the third term is a combination of two terms itself; it is clear that one part of this term is of order $O(\epsilon_f^0\epsilon_F)$ and the other part is of order $O(\epsilon_f\epsilon_F)$.
	
	In fact, \eqref{expanded distribu} can solve the kinetic equation up to first order in derivatives and in the small amplitude regime. We will show that by inserting it in the kinetic equation,  six equations can be found for exactly the six fields $\beta$, $\mu_R$, $\mu_L$ and $\boldsymbol{u}$, corresponding to six conservation laws in physical space.
	
	\subsection{Modifying the Momentum Current}
	As we advertised earlier, out of global equilibrium, the presence of magnetic field can modify the definition of the momentum current. Let us now see how it happens. 
	Lorentz invariance basically implies that \eqref{T0i} has to be equivalent with \eqref{Ti0}. We saw that this requirement leads to a modification in the of the energy dispersion in the equilibrium state \eqref{equi_dis_func} (see equation \eqref{epsilon_son}).
	We may reconsider the above Lorentz invariance condition
	when the system is out of global equilibrium. In this case, the second term in \eqref{T0i} will no longer vanish. So it is reasonable to demand that \eqref{Ti0} gets a correction as the following:
	\begin{equation}\label{Ti0_correct}
	T^{0i}=\int\frac{d^3\text{p}}{(2\pi)^3}\sqrt{G}\,\,\left(\text{p}_i+\frac{}{}f_i(\text{p})\right)\,n_{\mathbf{p}}
	\end{equation}
	where $f_i(\text{p})$ is a one derivative vector operator.  Using dispersion of energy in \eqref{epsilon_son},  the condition $T^{0i}=T^{i0}$ simplifies to the following equation, to first order in derivatives,
	\begin{equation}
-\int \frac{d^3 \text{p}}{(2\pi)^3}\epsilon^{ijk}\frac{\text{p}^2}{2}\,\lambda\,e\, \,\frac{\text{p}^j}{\text{p}^3}\frac{\partial n_{\textbf{p}}}{\partial \text{x}^k}=\int\frac{d^3\text{p}}{(2\pi)^3}\,\,\frac{}{}f_i(\text{p})\,n_{\mathbf{p}}.
	\end{equation} 
This equation shows that the momentum density \eqref{Ti0_correct} is given by 
\begin{equation}
\label{momentum_us}
\boldsymbol{\pi}=\int\frac{d^3\text{p}}{(2\pi)^3}\sqrt{G}\,\,\left(\textbf{p}-\frac{\lambda\, e}{2}\, \hat{\textbf{p}}\times \boldsymbol{\nabla}\right)\,n_{\mathbf{p}}.
\end{equation}
Clearly, the second contribution in the integrand does not survive in the equilibrium.
To our knowledge, this new contribution in the definition of momentum density in presence of magnetic field, was not well-known in the literature before. 
This modification compared to well known structure of momentum current, which is one of our results in this paper, suggests that in order to derive the equation of momentum conservation from the kinetic equation, the object which has to be multiplied with the kinetic equation is the following modified momentum operator
\begin{equation}\label{mom_modified}
\tilde{\textbf{p}}=\textbf{p}-\frac{\lambda\, e }{2}\, \hat{\textbf{p}}\times \boldsymbol{\nabla}.
\end{equation}
We will show that this is the right momentum which gives the hydrodynamic modes in the long wave-length limit of the chiral kinetic theory.

	\section{Perturbative Solution for the Kinetic Equation}
	\label{3}
	 
In the ansatz solution given by \eqref{expanded distribu}, there are six unknown hydro fields:  $\delta \beta$, $\boldsymbol{\pi}$ and $\delta \mu_{R,L}$.	
	At first sight it may seem that we want to find six unknown hydrodynamic fields from only one equation \eqref{kinetic_eq}. The presence of conservation laws
	in collisions however, allows us to derive exactly six equations for these six unknown fields.
It will be the main content of the following subsection.

	\subsection{Conservation Equations}
	 In order to derive the conservation equations from kinetic theory, 
	it is standard to  multiply the kinetic equation \eqref{kinetic_eq} by the collision invariant objects and then performing the integral over the momentum $\textbf{p}$. There are basically six conserved objects in the collisions in our system; the axial and vector charges, three components of  momentum and the energy. For each of these objects one may derive a macroscopic conservation equation from the kinetic equation.
	
The two equations associated with the axial and vector charge conservation in collisions may be alternatively given formally as
\begin{equation}\label{charge_cons}
\sum_{e}\int_{\textbf{p}}\,\,e\left(\frac{\partial n^{(\lambda,e)}_{\textbf{p}}(\textbf{x},t)}{\partial t}+\dot{\textbf{x}}\cdot\frac{\partial n^{(\lambda,e)}_{\textbf{p}}(\textbf{x},t)}{\partial \textbf{x}}+\dot{\textbf{p}}\cdot\frac{\partial n^{(\lambda,e)}_{\textbf{p}}(\textbf{x},t)}{\partial\textbf{p}}\right)=\sum_{e}\int_{\textbf{p}}\, \,e\,\,I_{coll}\{n^{(\lambda,e)}_{\textbf{p}}(\textbf{x},t)\}=0,\,\,\,\,\,\chi=R,L
\end{equation}
	where we have used the notation
$	\int_{\textbf{p}}\equiv\int \frac{d^3 \text{p}}{(2\pi)^3}\sqrt{G}$. Note that for each chirality we have to sum over the both particle ($e=1$) and anti-particle ($e=-1$) contributions. 
	In order to find the energy and momentum conservation equations, one multiples the kinetic equation \eqref{kinetic_eq} by $\epsilon({\textbf{p}})$ and  $\tilde{\textbf{p}}$ and integrate over $\textbf{p}$ and then sum over $e$ and $\lambda$: 
	\begin{eqnarray}\label{energy_cons}
	\sum_{e, \lambda}\int_{\textbf{p}}\epsilon(\textbf{p})\left(\frac{\partial n^{(\lambda,e)}_{\textbf{p}}(\textbf{x},t)}{\partial t}+\dot{\textbf{x}}\cdot\frac{\partial n^{(\lambda,e)}_{\textbf{p}}(\textbf{x},t)}{\partial \textbf{x}}+\dot{\textbf{p}}\cdot\frac{\partial n^{(\lambda,e)}_{\textbf{p}}(\textbf{x},t)}{\partial\textbf{p}}\right)&=&\sum_{e,\lambda}\int_{\textbf{p}}\epsilon(\textbf{p})\,\,I_{coll}\{n^{(\lambda,e)}_{\textbf{p}}(\textbf{x},t)\}=0,\\\label{momentum_cons}
		\sum_{e,\lambda}\int_{\textbf{p}}\tilde{\textbf{p}}\left(\frac{\partial n^{(\lambda,e)}_{\textbf{p}}(\textbf{x},t)}{\partial t}+\dot{\textbf{x}}\cdot\frac{\partial n^{(\lambda,e)}_{\textbf{p}}(\textbf{x},t)}{\partial \textbf{x}}+\dot{\textbf{p}}\cdot\frac{\partial n^{(\lambda,e)}_{\textbf{p}}(\textbf{x},t)}{\partial\textbf{p}}\right)&=&\sum_{e,\lambda}\int_{\textbf{p}}\tilde{\textbf{p}}\,\,I_{coll}\{n^{(\lambda,e)}_{\textbf{p}}(\textbf{x},t)\}=0.
	\end{eqnarray}
	These equations become exactly the hydrodynamic equations when considering in the long wave-length limit. To obtain the hydrodynamics at first order, we have to substitute \eqref{expanded distribu} in the above equations and keep terms up to second order in derivatives.  Let us recall that we would like to work in the regime in which the constitutive relations of hydrodynamics are up to first order in gradients and consequently, the equations of motion have to be taken to second order in gradients.   
	
	Let us take the hydro fields as $\phi_a=\{\delta\mu_{\chi}, \delta \beta, \boldsymbol{\pi} \},\,\,\,\,a=1,\cdots,6$.  In the following we work with the Fourier transformed of hydro fields, which is defined as 
	\begin{equation}
	\phi_a(\textbf{x},t)=\int d^3k\,d\omega\, e^{-i \omega t+ i \boldsymbol{k}\cdot\textbf{x}}\,\phi_a(\omega, \boldsymbol{k}).
	\end{equation}
	In this paper we always study the propagation in the direction of the magnetic field $\textbf{B}\parallel \boldsymbol{k}$ and so each $SO(3)$ vector like $\boldsymbol{\pi}$ would be given as 
	\begin{equation}
\boldsymbol{\pi}=(\pi^{\perp}_{1},\pi^{\perp}_{2},\pi^{\parallel})
	\end{equation}
	where $\pi^{\parallel}$ is the component of $\boldsymbol{\pi}$ parallel to the wave vector and $\pi^{\perp}_{1}$ and $\pi^{\perp}_{2}$ are the traverse  components of $\boldsymbol{\pi}$.  So the hydro fields may be simply given by
 \begin{equation}\label{variables}
 \delta \phi_a=\,(\delta \mu_{R,L}, \delta T, \pi^{\perp}_1\, \pi^{\perp}_2, \pi^{\parallel}), \,\,\, a=1,\cdots 6.
 \end{equation}
In the low amplitude regime, the six linearized conservation equations  may then written as
\begin{equation}\label{Mab}
M_{ab}(\omega,k)\,\delta \phi_{b}=\,0, \,\,\, a=1,\cdots 6.
\end{equation}
Our next task is to specify the elements of the matrix $M_{ab}(\omega,k)$. Let us denote that
 equations $a=1,2$ are the linearized form of the  \eqref{charge_cons}, the equations $a=3$ will be obtained by linearizing \eqref{energy_cons} and $a=4,5,6$ denote the equations \eqref{momentum_cons} when linearized.

To be more clear, we give in the following not only  the fully momentum integrated form of each equation, but as a priority,  we also give each equation before integrating.
In other words by considering
\begin{equation}
\int_{\textbf{p}}(\cdots)=\int\frac{d^3\text{p}}{(2\pi)^3}\sqrt{G}(\cdots)=\int \frac{\text{p}^2d\text{p}d\Omega}{(2\pi)^3}\sqrt{G}(\cdots)=\int \frac{d\text{p}\,\,d\Omega }{(2\pi)^{3}}\underbrace{\left(\text{p}^2 \sqrt{G}(\cdots)\right)}_{expanded\,\,\, integrand},
\end{equation}  
in the following subsection, we give the expanded integrands corresponding to hydro linearized equations together with the macroscopic hydro equation (see also Appendix \ref{App_4}).

\subsubsection{charge conservation}
In order to find the hydrodynamic equations corresponding to the charge conservation in collisions, let us consider equations \eqref{charge_cons}.
Using \eqref{expanded distribu}, we then can expand these equations to first order in hydro fields $\phi_a$  and formally write
	\begin{eqnarray} \label{current_eq}
  \sum_{e} \int_{\text{p}}\,  \,e\,  \mathcal{A}_{\delta \phi_{{a}}}^{\text{C}_{\chi}} \delta \phi_{a}=0,\,\,\,\,\,\,\,\chi=R,L
	\end{eqnarray}
where $\mathcal{A}_{\delta \phi_{{a}}}^{\text{C}_{\chi}}$ denotes the coefficient of $\delta \phi_a$ field in the equation of charge conservation (C) corresponding to the particles with $\chi$ chirality, before performing the integral over the momentum space \text{p}. There are  \textbf{twelve} of these coefficients for different values of  $a=1,\cdots,6$ and $\chi=R,L$. These  are given by
	\begin{eqnarray}\label{coeff_jR}
&& \mathcal{A}_{\delta \mu_{\chi'}}^{\text{C}_{\chi}}=\frac{\,i\, e \,\beta (\mathbf{p} \cdot \boldsymbol{k} - \text{p} \, \omega)\, e^{\beta(\text{p}-e\mu_{\chi})}}{\text{p}^{2} \left(e^{\beta(\text{p}-e\mu_{\chi})}+1\right)^{2}}\left\{\text{p}^{3} + \lambda \, \mathbf{B} \cdot \mathbf{p} \left[\frac{2\,\mathbf{p} \cdot \boldsymbol{k} - \text{p}\,\omega  }{\mathbf{p} \cdot \boldsymbol{k} -  \text{p}\,\omega} + \text{p}\, \beta \tanh\big(\frac{\beta}{2}(\text{p}-e\mu_{\chi})\big)\right]\right\},\,\,\,\,\,\,\,\,\,\,\,\,\,\,\,\chi, \chi'=\{R,L\}\nonumber\\
&&\mathcal{A}_{\delta \beta}^{\text{C}_{\chi}}=-\frac{\,i\, (\mathbf{p} \cdot \boldsymbol{k} -  \text{p}\,\omega) \,(\text{p} - e \mu_{\chi})\, e^{\beta(\text{p}-e\mu_{\chi})}}{\text{p}^{2} \left(e^{\beta(p-e\mu_{\chi})}+1\right)^{2}}\left\{\text{p}^{3} + \lambda \, \, \mathbf{B} \cdot \mathbf{p} \left[\frac{\mathbf{p} \cdot \boldsymbol{k}}{\mathbf{p} \cdot \boldsymbol{k} -  \text{p}\,\omega} - \frac{e \mu_{\chi}}{\text{p}-e\mu_{\chi}}+ \text{p}\, \beta \tanh\big(\frac{\beta}{2}(\text{p}-e\mu_{\chi})\big)\right]\right\},\nonumber\\
&&\mathcal{A}_{\pi^{\perp}_i}^{\text{C}_{\chi}}=\frac{ \beta \left(e (\mathbf{p}\times \mathbf{B})^{\perp}_{i} + i \mathbf{p}^{\perp}_{i}(\mathbf{p} \cdot \boldsymbol{k} - \omega \text{p})\right)  e^{\beta(\text{p}-e\mu_{\chi})}}{\text{p}^{2}\, w  \left(e^{\beta(\text{p}-e\mu_{\chi})}+1\right)^{2}}\nonumber\\
&& \hspace{1cm}\times \left\{\text{p}^{3} + \lambda \,\,  \left[\mathbf{B} \cdot \mathbf{p} \left(2+ \text{p} \, \beta \tanh\big(\frac{\beta}{2}(\text{p}-e\mu_{\chi})\big)\right)+ \frac{i \text{p}\, \mathbf{p}^{\perp}_{i}\left(\,\mathbf{B} \cdot \mathbf{p}\,\,\omega  - \frac{\text{p}}{2}\, \mathbf{B} \cdot \boldsymbol{k} \right)- e\,  \frac{\text{p}^{2}}{2} (\mathbf{p} \times \boldsymbol{k})^{\perp}_{i} (\mathbf{p} \cdot \boldsymbol{k} -  \text{p}\,\omega)}{ e(\mathbf{p}\times \mathbf{B})^{\perp}_{i} + i \mathbf{p}^{\perp}_{i}(\mathbf{p} \cdot \boldsymbol{k} -  \text{p}\,\omega)}\right] \right\},\nonumber\\
&&\,\,\,\,\,\,\,\,\,\,\,\,\,\,\,\,\,\,\,\,\,\,\,\,\,\,\,\,\,\,\,\,\,\,\,\,\,\,\,\,\,\,\,\,\,\,\,\,\,\,\,\,\,\,\,\,\,\,\,\,\,\,\,\,\,\,\,\,\,\,\,\,\,\,\,\,\,\,\,\,\,\,\,\,\,\,\,\,\,\,\,\,\,\,\,\,\,\,\,\,\,\,\,\,\,\,\,\,\,\,\,\,\,\,\,\,\,\,\,\,\,\,\,\,\,\,\,\,\,\,\,\,\,\,\,\,\,\,\,\,\,\,\,\,\,\,\,\,\,\,\,\,\,\,\,\,\,\,\,\,\,\,\,\,\,\,\,\,\,\,\,\,\,\,\,\,\,\,\,\,\,\,\,\,\,\,\,\,\,\,\,\,\,\,\,\,\,\,\,\,\,\,\,\,\,\,\,\,\,\,\,\,\,\,\,\,\,\,\,\,\,\,\,\,\,\,\,\,\,\,\,\,\,\,\,\,\,\,\,\,\,\,\,\,\,\,\,\,\,\,\,\,\,\,\,\,\,\,\,\,\,\,\,\,\,\,\,\,\,\,\,\,\,\,\,\,\,\,\,\,\,\,i=1,2\nonumber\\
&& \mathcal{A}_{ \pi^{\parallel}}^{\text{C}_{\chi}}=\frac{\,i \,\beta p^{\parallel} (\mathbf{p} \cdot \boldsymbol{k} -  \text{p}\,\omega)\, e^{\beta(\text{p}-e\mu_{\chi})}}{\text{p}^{2} w  \left(e^{\beta(\text{p}-e\mu_{\chi})}+1\right)^{2}}\left\{\text{p}^{3} + \lambda \,\,\mathbf{B} \cdot \mathbf{p} \left[\frac{2\,\mathbf{p} \cdot \boldsymbol{k} -  \text{p}\,\omega }{\mathbf{p} \cdot \boldsymbol{k} - \text{p}\,\omega } + \text{p}\, \beta \tanh\big(\frac{\beta}{2}(\text{p}-e\mu_{\chi})\big)\right]\right\}.
	\end{eqnarray}
 Let us mention that the last expression includes four of the twelve coefficients itself; $\mathcal{A}_{ \pi^{\parallel}_1}^{\text{C}_{R}}$, $\mathcal{A}_{ \pi^{\parallel}_2}^{\text{C}_{R}}$, $\mathcal{A}_{ \pi^{\parallel}_1}^{\text{C}_{L}}$ and $\mathcal{A}_{ \pi^{\parallel}_2}^{\text{C}_{L}}$. It should be also noted that the magnetic  field $\textbf{B}$ and wave vector $\boldsymbol{k}$ are assumed to be parallel in this paper. 
 Now by performing the integral over $\textbf{p}$, namely over $\text{p}$, $\theta$ and $\phi$ in spherical coordinates,  and doing sum over $e=\pm1$  in \eqref{current_eq}, and also by the use of equations \eqref{thermo},  we obtain the following two equations: 
 \begin{eqnarray}\label{current-right}
\big (	\bar{\boldsymbol{\chi}}_{R} \,\omega-\frac{\text{B} }{4 \pi^2}\,k\big)\,\delta \mu_R+\,\big(\mu_{R} \bar{\boldsymbol{\chi}}_{R} -3 n_{R}\big)\,\omega \,\, \frac{\delta\beta}{\beta}+\, \frac{\mu_{R}}{w}\left(\frac{\text{B} }{4 \pi^2}\,\omega-\frac{n_{R} }{\mu_R}\,k\right)\,\pi^{\parallel} &=&0,\\\label{current-left}
\big (	\bar{\boldsymbol{\chi}}_{L} \,\omega+ \frac{\text{B} }{4 \pi^2}\,k\big)\,\delta \mu_L+\,\big(\mu_{L} \bar{\boldsymbol{\chi}}_{L} -3 n_{L}\big)\,\omega \,\, \frac{\delta\beta}{\beta} - \, \frac{\mu_{L}}{w}\left(\frac{\text{B} }{4 \pi^2}\,\omega+ \frac{n_{L} }{\mu_L}\,k\right)\,\pi^{\parallel} &=&0.
 \end{eqnarray}
 These two equations are nothing but the equations of conservation for the right- and left-handed charge densities in the low energy to first order derivative regime. In the language of hydrodynamics, they are in correspondence with the linearized form of the following two hydrodynamic equations of the vector and axial currents
 \begin{eqnarray}
\partial_{\mu}J_V^{\mu}&=&0,\nonumber\\
\partial_{\mu}J_A^{\mu}&=& C E\cdot B,
\end{eqnarray} 
 where  vector  and axial currents are related to right- and left-handed charge currents through
$J_V^{\mu}=J^{\mu}_{R}+J^{\mu}_{L}$ and $J_A^{\mu}=J^{\mu}_{R}-J^{\mu}_{L}$.

 \subsubsection{energy conservation}
 In order to find the hydrodynamic equations corresponding to the energy conservation in collisions, let us consider equations \eqref{energy_cons}.
 Using \eqref{expanded distribu}, we then can expand these equations to first order in hydro fields $\phi_a$  and formally write

	 	\begin{eqnarray}\label{energy}
	    \sum_{\lambda} \sum_{e} \int_{\text{p}} \, \mathcal{A}_{\delta \phi_{{a}}}^{\text{E}} \delta \phi_{a}=0,
	 	\end{eqnarray}
	 	where $\mathcal{A}_{\delta \phi_{{a}}}^{\text{E}}$ denotes the coefficient of $\delta \phi_a$ field in the equation of energy conservation (E). The \textbf{six} $\mathcal{A}_{\delta \phi_{{a}}}^{\text{E}}$ coefficients with $a=1,\cdots,6$ are given by
	 	\begin{eqnarray}\label{coeff-energy}
	 	&&\mathcal{A}_{\delta\mu_{\chi}}^{\text{E}} = \text{p}\, \mathcal{A}_{\delta\mu_{{\chi}}}^{\text{C}_{\chi}} -  \frac{ i \, \lambda \, e\, \beta \, \mathbf{B} \cdot  \mathbf{p}\, (\mathbf{p} \cdot \boldsymbol{k} - \text{p}\,\omega ) e^{\beta(\text{p}-e\mu_{\chi})}}{\text{p}\, (e^{\beta(\text{p}-e\mu_{\chi})}+1)^{2}},\,\,\,\,\,\,\,\,\,\,\,\,\,\,\,\,\,\,\,\,\,\,\,\,\,\,\,\,\,\,\,\,\,\,\,\,\,\,\,\,\,\,\,\,\,\,\,\,\,\,\,\,\,\,\,\,\,\,\,\,\,\,\,\,\,\chi=R,L\nonumber\\
	 	&&\mathcal{A}_{\delta\beta}^{\text{E}}  = \text{p}\, \mathcal{A}_{\delta\beta}^{\text{C}_{\chi}} +  \frac{\, i \, \lambda \,  \, \mathbf{B} \cdot  \mathbf{p}\, (\text{p}- e \mu_{\chi})\, (\mathbf{p} \cdot \boldsymbol{k} - \text{p}\,\omega) e^{\beta(\text{p}-e\mu_{\chi})}}{ \text{p}\, (e^{\beta(\text{p}-e\mu_{\chi})}+1)^{2}},\nonumber\\
	 	&&\mathcal{A}_{\pi^{{\perp}}_i}^{\text{E}}  = \text{p}\, \mathcal{A}_{\pi^{\perp}_i}^{\text{C}_{\chi}} - \frac{\, \lambda \, \, \beta \,\mathbf{B} \cdot  \mathbf{p}\, e^{\beta(\text{p}-e\mu_{\chi})} \, \left( e\,(\mathbf{p}\times \mathbf{B})^{\perp}_i + i \mathbf{p}^{\perp}_i(\mathbf{p} \cdot \boldsymbol{k} - \text{p}\,\omega)\right)}{\text{p}\, w\,  (e^{\beta(\text{p}-e\mu_{\chi})}+1)^{2}},\,\,\,\,\,\,\,\,\,\,\,\,\,\,\,\,\,\,\,\,i=1,2\nonumber\\
	 	&&\mathcal{A}_{\pi^{\parallel}}^{\text{E}} = \text{p}\, \mathcal{A}_{\pi^{\parallel}}^{\text{C}_{\chi}} -  \frac{\, i \, \lambda\,\,  \beta \, p^{\parallel}\mathbf{B} \cdot  \mathbf{p}\, (\mathbf{p} \cdot \boldsymbol{k} - \, \text{p}\,\omega) e^{\beta(\text{p}-e\mu_{\chi})}}{ \text{p}\, w\,  (e^{\beta(\text{p}-e\mu_{\chi})}+1)^{2}}.
	 	\end{eqnarray}
	  Now one can perform the integral over $\textbf{p}$ and evaluate sum over $e=\pm1$ and $\lambda=\pm1/2$ in \eqref{energy}. Then by use of \eqref{thermo},  we obtain the following equation: 
	  \begin{equation}\label{en}
	  \left (	3 n_{R} \,\omega-\frac{\mu_{R}\,\text{B} }{4 \pi^2}\right)\,\delta \mu_R+\, \left(3 n_{L} \,\omega+\frac{\mu_{L}\,\text{B} }{4 \pi^2}\,k\right)\,\delta \mu_L- C_{v} \,\omega \,\, \frac{\delta\beta}{\beta^2}+\, \left(\frac{\bar{\boldsymbol{\chi}}_{R}-\bar{\boldsymbol{\chi}}_{L}}{2 w}\,\text{B}\, \omega-k \right)\,\pi^{\parallel} =0.
	  \end{equation}
	  This equation is the energy conservation equation in a system of right- and left-handed fermions in the long wavelength limit.
	   This equation is the linearized form of the following hydrodynamic equation: 
	  \begin{equation}
	  \partial_{\mu}T^{\mu0}=F^{0\mu}J_{\mu}.
	  \end{equation} 
Before leaving this subsection, let us give a comment concerning three equation  \eqref{current-right}, \eqref{current-left} and \eqref{en}. If one forces the momentum perturbations to be turned off, namely $\pi^{\parallel}=0$, these equations make a complete set of equations for the linear perturbations of temperature together with the right- and left-handed charge density perturbations. Such situation has been considered in \cite{Frenklakh:2016izv} (see also 	\cite{Chernodub:2015gxa}). In low density high temperature or inversely in low temperature high density limit, the temperature perturbations decouple as well. The latter case has been studied in 	\cite{Stephanov:2014dma}. In general, the cases studied in \cite{Frenklakh:2016izv,Stephanov:2014dma} are both related to forced fluids. In this paper however we let all six types of perturbations hydrodynamically propagate in the system. So to close the equations, we need to derive the three other equations associated with the momentum conservation in the system. ch computation would be the subject of next subsection.  	
	  \subsubsection{momentum conservation}
	  Let us now consider equation \eqref{momentum_cons}. Using \eqref{mom_modified} and also \eqref{expanded distribu}, we expand this equation and keep linear terms in fluctuations. To second order in derivative expansion we find  
	  \begin{eqnarray}\label{momentum}
	  \sum_{\lambda} \sum_{e} \int_{\text{p}} \, \mathcal{A}_{\delta \phi_{{a}}}^{\text{P}} \delta \phi_{a}=0,
	  \end{eqnarray}
	  where $\mathcal{A}_{\delta \phi_{{a}}}^{\text{P}}$ denotes the coefficient of $\delta \phi_a$ field in the equation of momentum conservation (P). There are two different group of these coefficients in \eqref{momentum}. Six of them are related to momentum conservation in the direction of wave vector. We show them by $\mathcal{A}_{\delta \phi_{{a}}}^{\text{P}^{\parallel}}$. The other group contains eighteen coefficients of momentum conservation in directions perpendicular to the wave propagation. We demonstrate them by $\mathcal{A}_{\delta \phi_{{a}}}^{\text{P}^{\perp}}$. Let us first give the coefficients of first group. It turns out that
	 		\begin{equation}\label{coeff-mompar}
	 		\mathcal{A}_{a}^{\text{P}^{{\parallel}}} = p^{\parallel} \mathcal{A}_{a}^{\text{C}_{\chi}}.
	 		\end{equation} 
	 		  Now by performing the integral over $\text{p}$ and doing sum over $e=\pm1$ and $\lambda=\pm1/2$ in \eqref{current_eq}, and also by use of \eqref{thermo},  we obtain the following equation: 
	 		\begin{equation}\label{mom-parallel}
	 		\left (		\frac{\mu_{R}}{4 \pi^2}\, \text{B} \,\omega-n_{R}\,k\right)\,\delta \mu_R-\, \left(\frac{\mu_{L} }{4 \pi^2}\, \text{B}\,\omega+n_{L}\,k\right)\,\delta \mu_L-\frac{ C_{v}}{3}\,k\,\frac{\delta\beta}{\beta^2}+\, \left(\omega- \frac{\chi_{R}-\chi_{L}}{2 w}\,\text{B}\, k \right)\,\pi^{\parallel} =0.
	 		\end{equation}
	 	The \textbf{twelve} $\mathcal{A}_{\delta \phi_{{a}}}^{\text{P}^{\perp}_i}$ coefficients with $a=1,\cdots,6$ and $j=1,2$ are given by
	 			\begin{eqnarray}\label{coeff-momperp}
	 			&&\mathcal{A}_{\delta\mu_{{\chi}}}^{\text{P}^{{\perp}}_{j}} = \mathbf{p}^{\perp}_j \mathcal{A}_{\delta\mu_{{\chi}}}^{\text{C}_{\chi}} +  \frac{ \lambda\, \, \beta (\mathbf{p}\times \boldsymbol{k})^{\perp}_j\, (\mathbf{p} \cdot \boldsymbol{k} - \text{p}\,\omega)\, e^{\beta(\text{p}-e\mu_{\chi})}}{2 (e^{\beta(\text{p}-e\mu_{c})}+1)^{2}},\,\,\,\,\,\,\,\,\,\,\,\,\,\,\,\,\,\,\,\,\,\,\,\,\,\,\,\,\,\,\,\,\,\,\,\,\,\,\,\,\,\,\,\,\,\,\,\,\,\,\,\,\,\,\,\,\,\,\,\,\,\,\,\,\,\,\,\,\,\,\,\,\,\,\,\,\,\,\,\chi=R,L\nonumber\\
	 				&&\mathcal{A}_{\delta\beta}^{\text{P}^{{\perp}}_{j}}  = \mathbf{p}_j^{\perp} \mathcal{A}_{\delta\beta}^{\text{C}_{\chi}} - \frac{ \lambda\, e\, (\mathbf{p}\times \boldsymbol{k})^{\perp}_j\, (\text{p}- e \,\mu_{\chi})\, (\mathbf{p} \cdot \boldsymbol{k} - \text{p}\,\omega)\, e^{\beta(\text{p}-e\mu_{\chi})}}{2 (e^{\beta(\text{p}-e\mu_{c})}+1)^{2}},\nonumber\\
	 					&&\mathcal{A}_{\pi^{{\perp}}_{i}}^{\text{P}^{{\perp}}_{j}}  = \mathbf{p}^{\perp}_j \mathcal{A}_{\pi^{\perp}_i}^{\text{C}_{\chi}} - \frac{i \lambda \,  e\, \beta (\mathbf{p}\times \boldsymbol{k})^{\perp}_j\, \, \left( e(\mathbf{p}\times \mathbf{B})_{i} + i\,  \mathbf{p}_{i}(\mathbf{p} \cdot \boldsymbol{k} - \omega \text{p}) \right)e^{\beta(\text{p}-e\mu_{\chi})}}{ 2 w\,  (e^{\beta(\text{p}-e\mu_{\chi})}+1)^{2}},\,\,\,\,\,\,\,\,\,\,\,\,\,\,\,\,\,\,\,\,\,\,\,\,\,\,\,\,\,\,i=1,2,\nonumber\\
	 						&&\mathcal{A}_{\pi^{\parallel}}^{\text{P}^{{\perp}}_{j}} = \mathbf{p}^{\perp}_j \mathcal{A}_{\pi^{\parallel}}^{\text{C}_{\chi}} + \frac{ \lambda\, e\, \beta\, p^{\parallel} (\mathbf{p}\times \boldsymbol{k})^{\perp}_j\, \, (\mathbf{p} \cdot \boldsymbol{k} - \text{p}\,\omega)\, e^{\beta(\text{p}-e\mu_{\chi})}}{2 w\,  (e^{\beta(\text{p}-e\mu_{\chi})}+1)^{2}}.
	 			\end{eqnarray}
	 	  As in the parallel case, by performing the integral over $\textbf{p}$ and doing sum over $e=\pm1$ and $\lambda=\pm1/2$ in \eqref{momentum}, and also by use of \eqref{thermo},  we obtain the following two equations: 
	 	\begin{eqnarray}\label{mom-perp1}
	 \omega\, \pi^{\perp}_1 -\frac{i}{w}\left((n_{R}+n_{L}) \text{B} +\frac{1}{2}\, k\,\omega\, (n_{R}- n_{L})\right) \pi^{\perp}_2 &=&0,\\\label{mom-perp2}
	 \frac{i}{w}\left((n_{R}+n_{L}) \text{B} +\frac{1}{2}\, k\,\omega\, (n_{R}- n_{L})\right) \pi^{\perp}_1+\,	 \omega\, \pi^{\perp}_2  &=&0.
	 	\end{eqnarray}
	 	The set of three equations \eqref{mom-parallel}, \eqref{mom-perp1} and \eqref{mom-perp2} are the linearized equations of conservation of momentum in a system of right- and left-handed fermions, in the long wavelength limit. These equations are in fact the linearized form of the following hydrodynamic equations:
	 	  \begin{equation}
	 	  \partial_{\mu}T^{\mu i}=F^{i \mu}J_{\mu}.
	 	  \end{equation} 
	 	Having specified the linearized equations perturbations in the long wavelength limit, our next task is to find the set of eigen modes of these equations. 
	 	\subsection{Hydro Modes}
	 	To find the hydrodynamic modes in our system, we now collect the linearized equations found in previous subsection together. Recalling the consised notation \eqref{Mab}, all the information of the equations (\ref{current-right}, \ref{current-left}, \ref{en}, \ref{mom-parallel}, \ref{mom-perp1} and \ref{mom-perp2}) may be simply given by the matrix $M_{ab}$. We may write
	 			 				\begin{equation}\label{MatrixB1}
	 				\resizebox{.99\hsize}{!}{$
	 		M_{ab}= 	\begin{bmatrix}
	 			\bar{\boldsymbol{\chi}}_{R} \,\omega-\frac{\text{B} }{4 \pi^2}\,k &0& T(\mu_{R} \bar{\boldsymbol{\chi}}_{R} -3 n_{R})\,\omega & 0 &0& \frac{\mu_{R}}{w}\left(\frac{\text{B} }{4 \pi^2}\,\omega-\frac{n_{R} }{\mu_R}\,k\right) \\
	 					0 &\bar{\boldsymbol{\chi}}_{L} \,\omega+\frac{\text{B} }{4 \pi^2}\,k& T(\mu_{L} \bar{\boldsymbol{\chi}}_{L} -3 n_{L})\,\omega & 0 &0& -\frac{\mu_{L}}{w}\left(\frac{\text{B} }{4 \pi^2}\,\omega+\frac{n_{L}}{\mu_L}\,k\right)\\
	 				3 n_{R} \,\omega-\frac{\mu_{R}\,\text{B} }{4 \pi^2}\,k&3 n_{L} \,\omega+\frac{\mu_{L}\,\text{B} }{4 \pi^2}\,k& -\omega C_{v} T^2 &  0 &0& \frac{(\chi_{R}-\chi_{L}) \,\text{B}}{2 w}\, \omega-k \\
	 				0 & 0 &0&\omega&-\frac{i}{w}\left((n_{R}+n_{L}) \text{B}+\frac{1}{2}\, k\,\omega\, (n_{R}- n_{L})\right)& 0\\
	 				0 & 0 &0&\frac{i}{w}\left((n_{R}+n_{L})\text{B}+\frac{1}{2}\, k\,\omega\, (n_{R}- n_{L})\right)&\omega& 0\\
	 				\frac{\mu_{R}\, \text{B} }{4 \pi^2}\,\omega-n_{R}\,k & -\frac{\mu_{L}\, \text{B} }{4 \pi^2}\,\omega-n_{L}\,k& -\frac{k C_{v}T^{2}}{3}&0& 0 &\omega- \frac{(\bar{\boldsymbol{\chi}}_{R}-\bar{\boldsymbol{\chi}}_{L}) \,\text{B}}{2 w}\, k
	 				\end{bmatrix}
	 				$}.
	 				\end{equation} 
	 			The six linear coupled algebraic equations have been replaced with the above matrix. In order to have non-trivial solution for the system of equations we require to have
	 			\begin{equation}\label{det}
	 			det(M)=\,0.
	 			\end{equation}
	 			When solving this equation in a power series in $k$, the corresponding roots, namely $\omega(k)$'s, specify the dispersion of hydrodynamic modes in the system. Since the linearized equations have been found to second order in derivatives in this paper, we are able to find the hydrodynamic modes to order $\epsilon_f^2$ as well.      To this end we set:
	 				 \begin{eqnarray}\label{derivative_expansion}
	 				 \omega \to \epsilon_f\, \omega^{(0)}+\epsilon_f^2\, \omega^{(1)},\,\,\,\,\,\,\,	\text{B}\to \epsilon_f \text{B},\,\,\,\,\,\,k\to \epsilon_f k,
	 				 \end{eqnarray}
	 				 where $\omega^{(0)}$ and $\omega^{(1)}$ correspond to dispersion at zero and first order hydrodynamics, respectively.
	 				    Using \eqref{derivative_expansion}, it turns out that the equation \eqref{det} contains terms which are at least of sixth order in gradients, i.e. $O(\epsilon_f^6)$:
	 				    \begin{equation}\label{det_expanded}
	 				    det(M)=\,\mathcal{E}_6\,\epsilon_f^6+\;\mathcal{E}_7\,\epsilon_f^7+\,O(\epsilon_f^8)=\,0.
	 				    \end{equation}
	 				   The leading order in the expression of hydro modes can be simply computed by solving  $\mathcal{E}_6=0$.  One obtains:
	 				    \begin{eqnarray}\label{zero_CMW}
\omega^{(0)}_{1,2}(k)&=&0,\\ \label{zero_sound}
\omega^{(0)}_{3,4}(k)&=&\pm\frac{1}{\sqrt{3}}k,\\ \label{zero_Alfven}
\omega^{(0)}_{5,6}(k)&=&\pm \frac{n_R+n_L}{w}\,\text{B}.
	 				    \end{eqnarray}
	 				    Substituting \eqref{zero_CMW} in \eqref{det_expanded} we obtain
	 				    \begin{equation}
det(M)=\left[\big(\omega^{(1)}_{1,2}\big)^{2}+\frac{\mathcal{A}_1}{2 \mathcal{E}}\textbf{B}\cdot \boldsymbol{k} \,\,\omega^{(1)}_{1,2}+ (\mathcal{A}_1^2-\mathcal{A}_2^2+4 \mathcal{A}_3\, \mathcal{E})\frac{\textbf{B}^2}{4 \mathcal{E}^2}\,\boldsymbol{k}^2\right]+\,O(\epsilon_f^9)=\,0
	 				    \end{equation}
	 				The coefficients have been defined in the following.     This gives 
	 				    \begin{equation}
\omega^{(1)}_{1,2}(k)=-\left(\mathcal{A}_1\pm\sqrt{\mathcal{A}_2^2- \,4\,\mathcal{A}_3\,\mathcal{E}} \right) \frac{1}{2\,\mathcal{E}}\,\text{B}\,k.
	 				    \end{equation}
	 				    Let us compute the correction  to the sound waves. Equation \eqref{det_expanded} for both modes in \eqref{zero_sound} simplifies to
	 				    \begin{equation}
	 				    det(M)=\left[(\bar{\boldsymbol{\chi}}_{R}-\bar{\boldsymbol{\chi}}_L) \textbf{B}\cdot \boldsymbol{k} - \frac{}{} 6 w \, \omega^{(1)}_{3,4}\right]+\,O(\epsilon_f^8)=\,0
	 				    \end{equation}
	 				    and so we have
	 				    \begin{equation}
	 				    \omega^{(1)}_{3,4}(k)=\frac{ \bar{\boldsymbol{\chi}}_{R}-\bar{\boldsymbol{\chi}}_L}{6 \, w}\,\text{B}\,k.
	 				    \end{equation}
	 				    Finally for both of modes given in \eqref{zero_Alfven} we find from \eqref{det_expanded} 
	 				     \begin{equation}
	 				    det(M)=\left[(n^2_{R}-n^2_L) \textbf{B}\cdot \boldsymbol{k} - \frac{}{} 2 w^2 \, \omega^{(1)}_{5,6}\right]+\,O(\epsilon_f^8)=\,0
	 				    \end{equation}
	 				    which leads to 
	 				    \begin{equation}
	 				    \omega^{(1)}_{5,6}(k)=\frac{(n_{R}+n_{L})(n_{R}-n_{L})}{2 w^2}\,\text{B}\,k.
	 				    \end{equation}
	 				   Collecting the zero order together with the first order parts of the modes, we may write the set of hydrodynamic modes up to first order on derivative expansion as it follows:
	 				\begin{eqnarray}\label{modes}\nonumber
	 			\omega_{1,2}(k)&=&-\left(\mathcal{A}_1\pm\sqrt{\mathcal{A}_2^2- \,4\,\mathcal{A}_3\,\mathcal{E}} \right) \frac{1}{2\,\mathcal{E}}\,\text{B}\,k,\\ 
	 			\omega_{3,4}(k)&=&\,\pm \frac{1}{\sqrt{3}}\,k+\,\frac{ \bar{\boldsymbol{\chi}}_{R}-\bar{\boldsymbol{\chi}}_L}{6 \, w}\,\text{B}\,k,\nonumber\\
	 			\omega_{5,6}(k)&=&\,\pm \frac{n_{R}+n_{L}}{w}\,\text{B}+\,\frac{(n_{R}+n_{L})(n_{R}-n_{L})}{2 w^2}\,\text{B}\,k,
	 				\end{eqnarray}
	 		where by use of the thermodynamic relations \eqref{thermo}  with the susceptibility coefficients (see Appendix \ref{suseptibility}) we may write	 				
	 				\begin{eqnarray}\label{anomaly_coef}
	 				&&\mathcal{E}=- \epsilon^{ijk} \alpha^{i} \beta^{j} \gamma^{k},~~ (\epsilon^{123}=1),\nonumber\\ \nonumber
	 			&&\mathcal{A}_{0} = \alpha_{{[1}} \gamma_{{2]}} \left(\frac{\left(n_{R} + n_{L}\right) \left(\mu_{R} + \mu_{L}\right)}{2} + \frac{\left(\mu_{R}^{2} - \mu_{L}^{2}\right)^{2}}{8\pi^{2}}\right) -  \alpha_{{[1}} \gamma_{{3]}} \left(\left(n_{R} + n_{L}\right) \left(\mu_{R} - \mu_{L}\right) + \frac{\left(\mu_{R}^{2} - \mu_{L}^{2}\right)\left(\mu_{R}+ \mu_{L}\right)^{2}}{24\pi^{2}}\right)\\
	 				&&\,\,\,\,\,\,\,\,\,\,\,\,\,\,\, + \alpha_{{[1}} \beta_{{3]}} \left(\frac{\left(n_{R} - n_{L}\right) \left(\mu_{R} - \mu_{L}\right)}{2} + \frac{\left(\mu_{R}^{2} - \mu_{L}^{2}\right)^{2}}{8\pi^{2}}\right)\nonumber -  \alpha_{{[1}} \beta_{{2]}} \left(\left(n_{R} - n_{L}\right) \left(\mu_{R} + \mu_{L}\right) + \frac{\left(\mu_{R}^{2} - \mu_{L}^{2}\right)\left(\mu_{R}- \mu_{L}\right)^{2}}{24\pi^{2}}\right)\nonumber\\
	 				&&\hspace{1.85cm}+\frac{T \left(\mu_{R}^{2} - \mu_{L}^{2}\right) \left(\alpha_{{[2}} \gamma_{{3]}} \left(\mu_{R}+ \mu_{L}\right) + \alpha_{{[3}} \beta_{{2]}} \left(\mu_{R}- \mu_{L}\right)\right)}{12},\nonumber\\
	 					&&\mathcal{A}_1=\frac{1}{4\pi^2}\left\{\alpha_{{[3}} \beta_{{1]}} + \alpha_{{[2}} \gamma_{{1]}}+ \frac{\mathcal{A}_{0} }{ w}\right\} ,\nonumber\\
	 					&&\mathcal{A}_2=\frac{1}{4\pi^2}\left\{\alpha_{{[3}} \beta_{{1]}} + \alpha_{{[2}} \gamma_{{1]}}+ \frac{2\mathcal{A}_{0} +\left(\mu_{R}^{2}-\mu_{L}^{2}\right) \mathcal{E}}{2 w}\right\} ,\nonumber\\
	 					&&\mathcal{A}_3= \frac{C_{v}^{2} T}{12 \pi^{4} w}.
	 				\end{eqnarray}
	 			Note that we used the convention $ 				A_{{[i}} B_{{j]}} \equiv A_{i} B_{j} - A_{j} B_{i}$.

In order to show that	each of the hydrodynamic modes \eqref{modes} carry which set of perturbations, we now compute the eigenvectors corresponding to these modes. To proceed we   reconsider equations \eqref{variables} and \eqref{Mab}. 				
The eigenvector corresponding to the $i^{th}$ mode is a vector in the six dimensional space of perturbations and we call it $\delta \phi_{a}^i$. It has to obey the following equation
\begin{equation}\label{eigenvector}
M_{ab}(\omega_i,k)\,\delta \phi_{b}^i=\,0, \,\,\, a=1,\cdots 6.
\end{equation}
For the first two modes in \eqref{modes} we obtain
\begin{equation}\label{eigen_vec_B_CMW}
\delta \phi^{1,2}=\,\left( -r,\, -s,  \,r\,\frac{\alpha_2}{\alpha_1}+s\,\frac{\alpha_3}{\alpha_1}, \,0,\, 0,\, 0\right),
\end{equation}
	with arbitrary parameters $r$ and $s$ (see \cite{Abbasi:2016rds} for more details). As it is seen, these modes carry the perturbations of right- and left-handed charge densities together with the temperature perturbation.  So these are actually the chiral magnetic heat waves.  These modes might be referred to as the scalar waves \cite{Yamamoto:2015ria,Abbasi:2016rds}. Forcing the momentum perturbations to be turned off,  the temperature perturbation decouples and we obtain the famous CMW \cite{Kharzeev:2010gd,Stephanov:2015roa}.  Let us consider the next two modes in \eqref{modes}. It turns out that 
	\begin{equation}\label{eigenvec34}
	\begin{split}
	\delta \phi^{3,4}=\left(\frac{C_1}{C_2},\frac{C_3}{C_2},\,1,\, 0,\, 0,\, \mp\frac{w \mathcal{E}}{C_2}c_s^2    \right)
	\end{split}
	\end{equation}
	with
	\begin{eqnarray}\label{C_1}
	C_1&=&n\, \alpha_{[1}\gamma_{3]} -n_5\, \alpha_{[1}\beta_{3]}-w\, \beta_{[1}\gamma_{3]} \\\label{C_2}
	C_2&=& n\, \alpha_{[3}\gamma_{2]}-n_5\, \alpha_{[3}\beta_{2]}-w\, \beta_{[3}\gamma_{2]} \\\label{C_3}
	C_3&=&  n\, \alpha_{[2}\gamma_{1]}-n_5\, \alpha_{[2}\beta_{1]}-w\, \beta_{[2}\gamma_{1]}.
	\end{eqnarray}
	These modes carry the longitudinal momentum perturbation; so they are the sound modes. However, since they carry the perturbations of right- and left-handed charge densities as well, they are in fact the modified sound modes. In the zero density limit where $n=n_5=0$, both $C_1$ and $C_3$ vanish and these modes become the ordinary sound waves.

The last two modes in \eqref{modes} correspond to the following eigenvectors:
\begin{equation}\label{eigenvec56}
\begin{split}
\delta \phi^{5,6}=\left(0,\frac{}{} 0,\, 0,\, 1,\,  \pm i,\, 0\right)
\end{split}
\end{equation}
These waves carry pure vector perturbations and we refer to them as the vector waves. These wave are the analogue of the chiral Alfv\'en waves
found in \cite{Yamamoto:2015ria,Abbasi:2015saa,Abbasi:2016rds}. 
	 				
It is worth mentioning that the hydrodynamic modes in a chiral fluid with right and left chiralities has been studied before in \cite{Abbasi:2016rds}, in the context of hydrodynamics.  It is interesting to compare the results obtained here from the chiral kinetic theory with those in \cite{Abbasi:2016rds}. The latter comparison would be the subject of next the sections in the current paper.

\section{Review of Hydro modes from Hydrodynamics in Landau-Lifshitz  Frame}
\label{4}

In the previous section, we computed the hydro modes in a chiral system of both right- and left-handed chiral fermions in presence of a homogeneous magnetic field. 
It is also possible  to compute the hydro modes of such system directly from the hydrodynamics. Such computations have been done in \cite{Abbasi:2015saa} for the case of fluid with single-chirality and also in  \cite{Abbasi:2016rds} for the same case that we are microscopically studying in the current paper.
In the following and for our later requirements, we briefly review the structure of chiral hydrodynamic equations and give the results obtained in  \cite{Abbasi:2016rds}.

The hydrodynamic equations are simply the  conservation equations. Getting $J^{\mu}_{V}$ and $J^{\mu}_{A}$  as the vector and axial currents, the hydro equations in presence of background magnetic field are
\begin{equation}\label{EoM}
\begin{split}
\partial_{\mu}T^{\mu \nu}=&\,F^{\nu \lambda} J_{V\lambda},\\
\partial_{\mu} J^{\mu}_{V}=&\, 0,\\
\partial_{\mu} J^{\mu}_{A}=&\, \mathcal{C} E_{\mu} B^{\mu},
\end{split}
\end{equation}
where we have defined the electric and magnetic field in the rest frame of this fluid  as $B^{\mu}=\frac{1}{2}\epsilon^{\mu\nu\alpha\beta}u_{\nu}F_{\alpha \beta}$ and  $E^{\mu}=\,F^{\mu \nu}u_{\nu}$, respectively.
The constitutive relations of energy-momentum and charge in a general hydrodynamic frame are 
\begin{equation}
\label{Tmunu_general}
\begin{split}
T^{\mu \nu}=& \,(\epsilon+p) u^{\mu} u^{\nu}+ p \,\eta^{\mu \nu} +\tau^{\mu \nu},\\
J^{\mu}_{V,A}=& \,n_{V,A}\,\, u^{\mu} +\nu^{\mu}_{V,A}.
\end{split}
\end{equation}
Here, $\tau^{\mu \nu}$ and $\nu^{\mu \nu}$ are the derivative corrections. Ignoring these corrections, \eqref{Tmunu_general} are showing the constitutive relations of an ideal fluid in terms of six thermodynamical fields $u^{\mu}(x)$, $T(x)$, $\mu_V(x)$ and $\mu_A(x)$. The system in this case is called to be in local equilibrium state.
Out of local equilibrium, these thermodynamic fields will no longer be well-defined. At every order in derivative expansion, to these terms, one may add the derivative corrections of the higher order and redefine them without changing the dynamics at the that order. This six-degree ambiguity reflects itself in the structure of constitutive relations. So it can be be fixed by demanding the derivative corrections in  \eqref{Tmunu_general} obey exactly six constraints. These constraints are referred to as the choice of frame in hydrodynamics. 

In the well-known Landau-Lifshitz frame one demands:
\begin{eqnarray}
u_{\mu}\,\tau^{\mu \nu}&=&0,\\
u_{\mu}\,\nu_{V,A} ^{\mu}&=&0.
\end{eqnarray}
Up to first order in derivative expansion, in the absence of dissipation and in presence of background weak vorticity and magnetic fields, the constitutive relations are then fixed as the following in the Landau-Lifshitz frame:
\begin{eqnarray}
\label{Tmunu_LL}
T^{\mu \nu} &= & w\, u^{\mu} u^{\nu} + p\, g^{\mu \nu},\\
\label{jmu_LL}
J_V^{\mu} & = & n_V\, u^{\mu} + \xi_V\, \omega^{\mu}+\xi_{VB}\, B^{\mu},\\\label{jmu5_LL}
J_A^{\mu} & = & n_A\,u^{\mu} + \xi_A \,\omega^{\mu}+\xi_{AB}\, B^{\mu},
\end{eqnarray}
with four unknown transport coefficients $\xi$'s in the axial and vector currents. In these expressions $\omega^{\mu}=\frac{1}{2}\epsilon^{\mu\nu\alpha\beta}u_{\nu}\partial_{\alpha}u_{\beta}$ is the local vorticity of fluid. These coefficients, which sit in front of parity odd terms, have been shown to be related to non-dissipative processes in chiral fluid \cite{Kharzeev:2011ds}.  Using this fact and demanding the second law of thermodynamics, one can specify the structure of these coefficients in terms of thermodynamic variables together with the underlying anomaly coefficients \cite{Son:2009tf,Bhattacharya:2011tra,Neiman:2010zi} (see also \cite{Abbasi:2016rds})
\begin{equation}
\begin{split}\label{transport_coef}
\xi_V&=2\mathcal{C}\, 
\left(\mu_V \mu_A-\frac{n_V \mu_A}{3w}\left(3\mu_V^2+\mu_A^2\right)\right)-2\mathcal{D}\,\frac{n_V \mu_A}{w}T^2,\\
\xi_A&=\mathcal{C}\, 
\left(\mu_V^2+ \mu_A^2-\frac{2 n_A \mu_A}{3 w}\left(3 \mu_V^2+\mu_A^2\right)\right)+\mathcal{D}\left(1-\frac{2 n_A \mu_A}{w}\right)T^2,\\
\xi_{VB}&=\mathcal{C}\, \mu_{A}\left(1-\frac{ n_V \mu_V}{w}\right),\\
\xi_{AB}&=\mathcal{C}\,\mu_V\left(1-\frac{ n_A \mu_A}{w}\right),
\end{split}
\end{equation}

where $\mathcal{C}$ and $\mathcal{D}$ are the coefficients of chiral anomaly and gravitational anomaly respectively, as 
\begin{equation}
\mathcal{C}=\frac{1}{2 \pi^2},\,\,\,\,\,\,\,\,\,\,\mathcal{D}=\frac{1}{6}.
\end{equation}
In \cite{Abbasi:2016rds}, the long wave-length excitations of the above-mentioned chiral fluid have been computed in three different cases. From those, we are interested just in the case that the chiral fluid is coupled to an external constant magnetic field, in this paper. The equilibrium in the system is given by
\begin{equation}
u^{\mu}=\left(1,\frac{}{}\boldsymbol{0}\right),\,\,\,\,
\,\,\,\, T=Const.,\,\,\,\,\mu_V=Const.,\,\,\,\,\mu_A=Const.
\end{equation}
The hydro modes in the system are the long wave-length solutions of the equations \eqref{EoM}, when linearized around the above equilibrium state. To linearize the equations, we consider the following perturbations
\begin{equation}
\phi_{a}+\delta \phi_a=\left(\mu_V+\delta \mu_V,\,\mu_A+\delta \mu_A,\,T+\delta T, \,\frac{}{}0+\pi^{\parallel}, \,\frac{}{}0+\pi^{\perp}_{1}, \,\frac{}{}0+\pi^{\perp}_{2}\right).
\end{equation}
The six coupled linear equations then may be formally written as 
\begin{equation}\label{linear_EOM}
M^{\boldsymbol{B}}_{ab}(\boldsymbol{k} , \omega)  \delta \phi_{ab} (\boldsymbol{k} , \omega) = 0,
\end{equation}
with
\begin{equation*}\label{MatrixB}
\begin{bmatrix}
-i \beta_1 \omega+\left(\frac{\partial \xi_{VB}}{\partial T}\right)  i\, \text{B} \,k  & -i\beta_2\omega +  \left(\frac{\partial \xi_{VB}}{\partial \mu_V}\right) i\, \text{B} \,k & -i \beta_3 \omega+ \left(\frac{\partial \xi_{VB}}{\partial \mu_A}\right) i\, \text{B} \,k& \frac{n_V}{w} i k - \frac{\xi_{VB}}{w} i\omega \text{B} &0 &0\\
-i \gamma_1 \omega+\left(\frac{\partial \xi_{A B}}{\partial T}\right) i\, \text{B} \,k  &  -i \gamma_2 \omega+ \left(\frac{\partial \xi_{A B}}{\partial \mu_V}\right) i\, \text{B} \,k & - i\gamma_3  \omega+ \left(\frac{\partial \xi_{A B}}{\partial \mu_A}\right) i\, \text{B} \,k&\frac{n_A}{w} i k - \frac{\xi_{AB}}{w} i\omega \text{B}&0 &0\\
- i \alpha_1\omega  & -i \alpha_2 \omega & -i \alpha_3 \omega & i k&0 &0\\
i\alpha_1  v_s^2 \,k  &  i\alpha_2  v_s^2 \,k & i\alpha_3  v_s^2 \,k & -i\omega & 0&0\\	
0 &0&0 & 0&-i \omega-\frac{\xi_V}{2 w}i \, \text{B}\,k &-\frac{n_V}{w}\text{B} \\
0&0 &0& 0& \frac{n_V}{w}\text{B} &-i \omega-\frac{\xi_V}{2 w}i \, \text{B}\,k
\end{bmatrix}.
\end{equation*}
By the same procedure used below \eqref{det}, one can find the hydrodynamic modes of the equations  \eqref{linear_EOM}.
It turns out that
\begin{eqnarray}\nonumber\label{modes_hydro}
	\omega_{1,2}(k)&=&-\left(\text{A}_1\pm\sqrt{\text{A}_2^2- \,4\,\text{A}_3\,\mathcal{E}} \right) \frac{1}{2\,\mathcal{E}}\,\text{B}\,k,\\ 
\omega_{3,4}(k)&=&\,\pm \frac{1}{\sqrt{3}}\,k,\\ \nonumber
\omega_{5,6}(k)&=&\,\pm \frac{n_V}{w}\,\text{B} -\frac{\xi_V }{2 w}\,\text{B}\,k,
\end{eqnarray}
with
\begin{eqnarray}
\text{A}_1=\text{A}_2&=&\alpha_{[2},\beta_{1]}\frac{\partial \xi_{A B}}{\partial \mu_{A}}-\alpha_{[2},\gamma_{1]}\frac{\partial \xi_{V B}}{\partial \mu_{A}}-\alpha_{[3},\beta_{1]}\frac{\partial \xi_{A B}}{\partial \mu_{V}}+\alpha_{[3},\gamma_{1]}\frac{\partial \xi_{V B}}{\partial \mu_{V}}+\alpha_{[3},\beta_{2]}\frac{\partial \xi_{A B}}{\partial T}-\alpha_{[3},\gamma_{2]}\frac{\partial \xi_{V B}}{\partial T},\\
\text{A}_3&=&\frac{\partial \xi_{V B}}{\partial \mu_{A}}\big(-\alpha_1  \frac{\partial \xi_{A B}}{\partial \mu_{V}}  +\alpha_2\frac{\partial \xi_{A B}}{\partial T}\big)+\frac{\partial \xi_{V B}}{\partial \mu_{V}}\big(\alpha_1  \frac{\partial \xi_{A B}}{\partial \mu_{A}}  -\alpha_3\frac{\partial \xi_{A B}}{\partial T}\big)+\frac{\partial \xi_{V B}}{\partial T}\big(\alpha_3  \frac{\partial \xi_{A B}}{\partial \mu_{V}}  -\alpha_2\frac{\partial \xi_{A B}}{\partial \mu_A}\big),\\
\mathcal{E}&=&-\epsilon^{ijk} \alpha_i \beta_j \gamma_k\,\,\,\,\,\,\,\,\,(\epsilon^{123}=1).
\end{eqnarray}
As discussed in detail in \cite{Abbasi:2016rds}, the first two modes, namely $\omega_{1,2}$, are the scalar waves by the mean that they carry the perturbations of vector and axial chemical potential together with the temperature perturbation. They are called Chiral-Magnetic-heat Waves (CMHW). The next four modes in general, when considering the propagation in an arbitrary direction with respect to the external magnetic field,  are mixed scalar-vector waves and may carry the perturbations of all six hydro fields. However when $\boldsymbol{B}\parallel \boldsymbol{k}$, which is assumed in this note, the scalar-vector modes are distinguished from each other. They become two
sound modes $\omega_{3,4}$ together with two chiral modes $\omega_{5,6}$ and propagate independently. The last two modes, namely $\omega_{5,6}$, which are excited due to coupling between vorticity and the magnetic field \cite{Abbasi:2016rds}, are referred to as the    Chiral-Alfv\'en waves (CAW)  \cite{Yamamoto:2015ria}.

In order to compare \eqref{modes} with \eqref{modes_hydro}, we use \eqref{transport_coef} to rewrite the modes \eqref{modes_hydro} as it follows
\begin{eqnarray}\nonumber\label{modes_hydro2}
\text{CMHW}\,\,\,\,\,\,\,\,\,\,\,\,\,\,\,\,\,\,\,\,	\omega_{1,2}(k)&=&-\left(\text{A}_1\pm\sqrt{\text{A}_2^2- \,4\,\text{A}_3\,\mathcal{E}} \right) \frac{1}{2\,\mathcal{E}}\,\text{B}\,k,\\ 
\text{Sound Waves}\,\,\,\,\,\,\,\,\,\,\,\,\,\,\,\,\,\omega_{3,4}(k)&=&\,\pm \frac{1}{\sqrt{3}}\,k,\\ \nonumber
\text{CAWs}\,\,\,\,\,\,\,\,\,\,\,\,\,\,\,\,\,\omega_{5,6}(k)&=&\,\pm \frac{n_R+n_L}{w}\,\text{B}+\left(\frac{(n_R-n_L)(n_R+n_L) }{2 w^2}-\frac{\bar{\boldsymbol{\chi}}_{R}-\bar{\boldsymbol{\chi}}_L}{4w}\right)\,\text{B}\,k,
\end{eqnarray}
		\begin{equation}
\text{A}_1=\text{A}_2=\mathcal{A}_2,\,\,\,\,\,\,\,\,
\text{A}_3=\mathcal{A}_3.
\end{equation}
with $\mathcal{A}_2$ and $\mathcal{A}_3$ defined in \eqref{anomaly_coef}.

In next section, we will explain how the hydrodynamic modes computed in the framework of chiral kinetic theory can be related to those found from the hydrodynamic in the Landau-Lifshitz frame.

\section{Modes in Lab Frame From those in Lanadau-Lifshitz Frame}
\label{sec_5}
Firstly let us recall that in the Landau-Lifshitz frame (LL) no energy flow would exist in the local rest frame of the fluid. This can be simply seen in \eqref{Tmunu_LL} when $u^{\mu}=(1,0,0,0)$:
\begin{equation}\label{T_0i_LL}
T_{LL}^{i0}=0.
\end{equation}

Now, we compute the same quantity in the long wave-length regime of CKT. Since the theory of Weyl fermions is Lorentz invariant, namely $T^{0i}=T^{i0}$, we can alternatively compute the momentum density  $T^{0i}$ as the energy flow in the local rest frame of the fluid. Using equilibrium distribution function \eqref{dis_func_thermo}, the equilibrium momentum density  \eqref{Ti0} reads, up to first order in derivative expansion 
\begin{equation}\label{T_0i_CKT}
\begin{split}
T_{CKT}^{i0}= T_{CKT}^{0i}=\pi_{CKT}^{i}=\,&\sum_{\lambda}\sum_{e}\int\frac{d^3\text{p}}{(2\pi)^3}\sqrt{G}\,\,\text{p}^i\,\left(	\tilde{n}^{(\lambda,e)}_{\textbf{p}}-\left(\frac{\partial 	\tilde{n}^{(\lambda,e)}_{\textbf{p}}}{\partial \epsilon(\textbf{p})}\right)_{eq.}\left(e \lambda\frac{\textbf{B}\cdot\textbf{p}}{\text{p}^2}\right) \epsilon_{f}\right)\\
=&\sum_{\lambda}\sum_{e}\int\frac{d^3\text{p}}{(2\pi)^3}\,\,e \lambda\,\, \left(\tilde{n}^{(\lambda,e)}_{\textbf{p}}-\big(\tilde{n}^{(\lambda,e)}_{\textbf{p}}\big)^2\right)\,\beta\,\text{B}_j\frac{\text{p}^i\text{p}^j}
{\text{p}^2}\\
=&\sum_{\lambda}\sum_{e}\int\frac{d\text{p}}{2\pi^2}\,\,e \lambda\,\, \left(\tilde{n}^{(\lambda,e)}_{\textbf{p}}-\big(\tilde{n}^{(\lambda,e)}_{\textbf{p}}\big)^2\right)\,\frac{1}
{3}
\beta\,\text{B}_i\\
=&\left(\frac{\mu_R^2-\mu_L^2}{8 \pi^2}\right)\,\text{B}_i=\,\,\frac{\bar{\boldsymbol{\chi}}_{R}-\bar{\boldsymbol{\chi}}_{L}}{4}\,\text{B}_i
\end{split}
\end{equation}
where from the second line to third one, we have exploited the isotropy of thermal state to replace $\text{p}^i\text{p}^j$ with $\frac{1}{3}p^2\delta^{ij}$ in the integrand.  We have also used the definition 
of the 	$\bar{\boldsymbol{\chi}}_{R,L}$ from \eqref{thermo}, in the last line.  This results had been found in the framework of quantum kinetic theory before \cite{Gao:2012ix}. 

Let us remember our convention about the momentum density, i.e. $\pi_i= v_{i}\,w$. Now,  considering \eqref{T_0i_LL} and \eqref{T_0i_CKT}, we   conclude that $T^{i0}_{LL}$ is computed in a Lorentz frame which is boosted with respect to the frame in which $T^{i0}_{CKT}$ is computed.
Let us emphasize that both \eqref{T_0i_LL} and \eqref{T_0i_CKT} are computed in the rest frame of the fluid in equilibrium. Consequently,\textbf{ in the equilibrium, the rest frames of the fluid in the two above frames are related to each other by the following boost} 
\begin{equation}\label{v_boost}
\boldsymbol{\beta}=\frac{\bar{\boldsymbol{\chi}}_{R}-\bar{\boldsymbol{\chi}}_{L}}{4 w}\,\textbf{B}.
\end{equation}
This idea, namely connecting the rest frames of the fluid in two different frames, has an important outcome. Since the hydrodynamic modes propagate in the rest frame on top of the equilibrium state in fluid,  we must be able to rederive the velocity of hydrodynamic modes in LL by applying the above Lorentz boost to those obtained from CKT.   

To see more clearly the difference between the two sets of results, in Table \ref{comparison}, we have shown the velocity of all six hydrodynamic modes in both cases.	
\begin{table}[!htb]
	\label{table one}
	\begin{center}
		\begin{tabular}{|c|c|c|}
			\hline
			\hline
			Type of mode &  Chiral Kinetic Theory & Landau-Lifshitz \\
			\hline
			\hline
			&  & \\ 
			CMHW    & $v^{CKT}_{1,2}(k)=-\left(\mathcal{A}_1\pm\sqrt{\mathcal{A}_2^2- \,4\,\mathcal{A}_3\,\mathcal{E}} \right)\frac{1}{2\,\mathcal{E}}\,B$ &  $v^{LL}_{1,2}(k)=-\left(\text{A}_1\pm\sqrt{\text{A}_2^2- \,4\,\text{A}_3\,\mathcal{E}} \right) \frac{1}{2\,\mathcal{E}}\,B$   \\
					
&   &  \\
\hline
&  & \\ 		
		Sound     & $v^{CKT}_{3,4}=\pm \frac{1}{\sqrt{3}}+\,\frac{\bar{\boldsymbol{\chi}}_{R}-\bar{\boldsymbol{\chi}}_{L}}{6 \, w}\,B$ &  $v^{LL}_{3,4}=\pm \frac{1}{\sqrt{3}}$  \\
		
		&   &  \\
		\hline
		&  & \\ 
		
		CAW     & $v^{CKT}_{5,6}=\,\frac{(n_{R}+n_{L})(n_{R}-n_{L})}{2 w^2}\,B$ &  $v^{LL}_{4,5}=\left(\frac{(n_R-n_L)(n_R+n_L) }{2 w^2}-\frac{\bar{\boldsymbol{\chi}}_{R}-\bar{\boldsymbol{\chi}}_{L}}{4w}\right)\,B$  \\
		
		&   &  \\
		\hline
		\hline
	\end{tabular}
\end{center}
\caption{\,\,\,\,\,\,\,\,\,\,\,$\text{A}_1=\mathcal{A}_1+\frac{\bar{\boldsymbol{\chi}}_{R}-\bar{\boldsymbol{\chi}}_{L}}{8\,w}\,\mathcal{E}$, $\text{A}_2=\mathcal{A}_2$ and $\text{A}_3=\mathcal{A}_3$.}
\label{comparison}
\end{table}
Obviously, the velocity of hydro modes computed in LL frame differ from those obtained from the hydro regime of CKT. To check the idea of boost, introduced above, let us compute the velocity of $v^{CKT}_i$ modes after making the boost \eqref{v_boost}:
\begin{equation}\label{boost}
v^{CKT}_{i}\,\,\,\,\rightarrow\,\,\,\,\frac{v^{CKT}_{i}-\beta}{1-v^{CKT}_{i} \beta   }.
\end{equation}
In the following we perform \eqref{boost} for each of $v_{i}^{CKT}$'s, up to first order in derivative expansion. To keep track of derivative corrections, we write each one derivative term with a $\epsilon_f$ factor and after expanding to first order in $\epsilon_f$, we set it to unity. In such counting, the velocity of boost given by \eqref{v_boost} is of order $\epsilon_f$ itself.

$	\bullet$ \textbf{Chiral Magnetic Heat Waves:}
\newline
Let us start with transforming the CMHWs. As mentioned earlier, these modes appear from the first order in derivative expansion. So we may write:
\begin{equation}\label{boost_12}
\begin{split}
v_{1,2}^{CKT}\,\,\,\,\rightarrow\,\,\,\,&\frac{-\left(\mathcal{A}_1\pm\sqrt{\mathcal{A}_2^2- \,4\,\mathcal{A}_3\,\mathcal{E}} \right)\frac{1}{2\,\mathcal{E}}\,\text{B}\,\epsilon_f-\,\frac{\bar{\boldsymbol{\chi}}_{R}-\bar{\boldsymbol{\chi}}_{L}}{4 w}\,\text{B}\,\epsilon_f}{1-\,O(\epsilon_f)^2}\\
&=-\left(\big(\mathcal{A}_1+\,\frac{\bar{\boldsymbol{\chi}}_{R}-\bar{\boldsymbol{\chi}}_{L}}{8 w}\,\mathcal{E}\,\text{B}\big)\pm\sqrt{\mathcal{A}_2^2- \,4\,\mathcal{A}_3\,\mathcal{E}} \right)\frac{1}{2\,\mathcal{E}}\,\text{B}\,\epsilon_f+\,O(\epsilon_f^2)\\
&=-\left(\text{A}_1\pm\sqrt{\text{A}_2^2- \,4\,\text{A}_3\,\mathcal{E}} \right) \frac{1}{2\,\mathcal{E}}\,\text{B}\, \epsilon_f\\
&=\,\,\boxed{v_{1,2}^{LL}}.
\end{split}
\end{equation}
From the second line to third line we have used the relation previously mentioned in the caption of Table \ref{comparison}.
As the first check of our idea, we derived the CMHWs in LL frame by boosting the velocity of CMHWs obtained from CKT. 

$	\bullet$ \textbf{Sound Waves:}
\newline
Since the sound velocity does exist in zeroth order, we should perform the computations more carefully:
\begin{equation}\label{boost_34}
\begin{split}
v_{3,4}^{CKT}\,\,\,\,\rightarrow\,\,\,\,&\frac{\pm\frac{1}{\sqrt{3}}+\,\frac{ \bar{\boldsymbol{\chi}}_{R}-\bar{\boldsymbol{\chi}}_{L}}{6 \, w}\,\text{B}\,\epsilon_f-\,\frac{\bar{\boldsymbol{\chi}}_{R}-\bar{\boldsymbol{\chi}}_{L}}{4 w}\,\text{B}\,\epsilon_f}{1-\left(\pm\frac{1}{\sqrt{3}}+\,\frac{ \bar{\boldsymbol{\chi}}_{R}-\bar{\boldsymbol{\chi}}_{L}}{6 \, w}\,\text{B}\,\epsilon_f\right)\frac{\bar{\boldsymbol{\chi}}_{R}-\bar{\boldsymbol{\chi}}_{L}}{4 w}\,\text{B}\,\epsilon_f}\\
&=\pm\frac{1}{\sqrt{3}}+\,\left(\big(\pm\frac{1}{\sqrt{3}}\big)^2\big(\frac{1}{4}\big)+\frac{1}{6}-\frac{1}{4}\right)\frac{ \bar{\boldsymbol{\chi}}_{R}-\bar{\boldsymbol{\chi}}_{L}}{ \, w}\,\text{B}\,\epsilon_f\\
&=\pm\frac{1}{\sqrt{3}}\\
&=\,\,\boxed{v_{3,4}^{LL}}.
\end{split}
\end{equation}
This computation confirms that the velocity of sound waves in LL frame gets no correction from the magnetic field.

$	\bullet$ \textbf{Chiral Alfv\'en Waves:}
\newline
Analogous to CMHWs, CAWs appear from the first order in derivative expansion. So we may write 
\begin{equation}\label{boost_56}
\begin{split}
v_{5,6}^{CKT}\,\,\,\,\rightarrow\,\,\,\,&\frac{\frac{(n_{R}+n_{L})(n_{R}-n_{L})}{2 w^2}\,\text{B}\,\epsilon_f-\,\frac{\bar{\boldsymbol{\chi}}_{R}-\bar{\boldsymbol{\chi}}_{L}}{4 w}\,\text{B}\,\epsilon_f}{1-\,O(\epsilon_f)^2}\\
&=\left(\frac{(n_R-n_L)(n_R+n_L) }{2 w^2}-\frac{\bar{\boldsymbol{\chi}}_{R}-\bar{\boldsymbol{\chi}}_{L}}{4w}\right)\,\text{B}\,\epsilon_f\\
&=\,\,\boxed{v_{5,6}^{LL}}.
\end{split}
\end{equation}

What we have shown above may be reviewed as it follows. In the long wave-length regime of the CKT, energy flows in the rest frame of the system like \eqref{T_0i_CKT}. On the other hand, relativistic hydrodynamic in the LL frame demands no energy flow exists in the rest frame of the fluid.
This difference between energy flows in equilibrium is the key point which motivates to relate two sets of results given in Table \ref{comparison} by Lorentz transformations to each other. As explicitly shown in \eqref{boost_12}, \eqref{boost_34} and \eqref{boost_56},
the relation is nothing but a Lorentz boost with the velocity \eqref{v_boost}.
Interestingly, \eqref{boost_12} is related to the shift in the frequency of sound due to \textbf{Doppler effect}.

 \textbf{Precisely speaking, the presence of such relation between the modes of two frames is the consequence of the fact that in equilibrium,  one can make the boost \eqref{v_boost} to transform the rest frame of the fluid in LL frame  to the rest frame of the chiral system in the kinetic theory side.}

Before ending this section let us mention that in a more general case we could consider the system not only being coupled to the magnetic field $\textbf{B}$, but also rotating around a fixed axis with frequency $\boldsymbol{\Omega}$.
As it was shown in \cite{Vilenkin:1979ui} for the CKT with $\textbf{B}=0$ and $\boldsymbol{\Omega\ne0}$,  the energy flow in the rest frame of the fluid might be non-zero due to global rotation in this case. It has been shown that in  the same situation, for the fluid with both right- and left-handed chiralities  \cite{Chen:2015gta}
\begin{equation}\label{T_0i_CKT_omega}
T^{i0}_{CKT}=T_{CKT}^{0i}=\pi_{CKT}^{i}=\,(n_R-n_L)\,\Omega_i.
\end{equation} 
This motivates to generalize the idea of boost \eqref{v_boost} to the case in which  a non-zero vorticity is present as well. So one expects the following boost to perform the desired Lorentz transformations in this case: 
\begin{equation}\label{general_boost}
\boldsymbol{\beta}=\frac{\bar{\boldsymbol{\chi}}_{R}-\bar{\boldsymbol{\chi}}_{L}}{4 w}\,\textbf{B}+\,\frac{(n_R-n_L)}{w}\,\boldsymbol{\Omega}.
\end{equation}

	\section{Lab Frame is a No-drag Frame}
	\label{sec_6}
	The hydrodynamic frame corresponding to the long wave-length limit of the kinetic  theory is the Laboratory (Lab) frame. In the Lab frame the velocity of fluid in equilibrium coincides with the four velocity of laboratory, to all orders in derivative expansion \cite{Jensen:2012jh}. 
	Our result in previous section so means that in the absence of dissipation one can transform the rest frame of the fluid in Landau-Lifshitz frame to that in Lab farme, by the boost \eqref{general_boost}.

There are two important points about the hydrodynamics in the Lab frame. First, the transport coefficients in this frame have been computed in different frameworks; from quantum kinetic theory \cite{Gao:2012ix}, microscopic thermal field theory \cite{Landsteiner:2012kd} and also from the lowest Landau level currents \cite{Landsteiner:2016led}. The second point is that the Lab frame is in fact a no-drag frame.
	In \cite{Stephanov:2015roa}, it has been shown that 
	in an anomalous fluid, 
	there is a frame in which a stationary obstacle experiences no drag. One can say that an impurity
    started to move under the influence of the drag force,	will be finally carried by the flow and it defines the
	“no-drag frame”. It can be shown that the constitutive relations in the no-drag frame is a general case of the constitutive relations in the  Lab frame \cite{Landsteiner:2016led}. 
	The computations of \cite{Stephanov:2015roa} are limited to the simple case of  an anomalous fluid with only one flavor.

	We generalize the computations of \cite{Stephanov:2015roa} to a system of non-Abelian chiral fermions and compute the transport coefficients in the no-drag frame for such system. Then we show that our results in a special case reduce to the constitutive relations for a system  of right- and left-handed fermions in the Lab frame. This simply says that the hydro limit of kinetic theory corresponds to a no-drag frame. To reconfirm the idea of boost, mentioned in previous section, we compute the velocity of boost, i.e. $T^{0i}_{no-drag}/w$, and boost our result to the LL frame. It turns out that the transport coefficients obtained from this transformations are exactly those obtained from hydrodynamics in a system with non-Abelian currents in LL frame \cite{Neiman:2010zi}.

	\subsection{No-Drag Frame for a System of Non-Abelian Fermions}

	Explicitly speaking,  we will find the  structure of the constitutive relations such that the drag force exerted on a quark in the fluid  gets  correction neither from the magnetic field nor from vorticity.   Since  it turns out that this drag force would no longer act on a rest quark in the rest frame of the fluid, the corresponding fluid frame would be a no-drag frame.

	Let us consider a point-like heavy quark in the fluid. Due to interaction with  the fluid, quark perturbs its surrounding medium and fluid exerts a friction-like force on the quark, the so-called drag force 	\cite{Chesler:2007sv}.  The drag force is minus the rate at which the quark deposits its momentum into the fluid.   In the absence of any external source, the quark gradually loses its energy and momentum and slows down. By applying an external force density  $\mathcal{F}^{\mu}$, e.g. an electric force,  we may supply the energy and momentum which  quark needs to move uniformly in the fluid. The conservation equations in this case may be written as  
	\begin{eqnarray}
		\partial_{\mu}T^{\mu \nu} &=& F^{ \nu \lambda}_a  J_{\lambda}^a+\,\mathcal{F}^{\nu},\\
		\partial_{\mu} J^{a \mu }& =&C^{abc} E^b\cdot B^c,
	\end{eqnarray} 
	where $\mathcal{F}^{\nu}$  is the local 4-momentum transfer from the quark to the fluid,
	i.e., the minus of drag force, per unit volume. In the above expressions, $C^{abc}$ refers to the
	anomaly coefficients and is totally symmetric
	under permutation of indices. We have also defined the field strength as 
	\begin{eqnarray}
	F^{a\lambda \nu} = E^{a\lambda} u^{\nu} - E^{a\nu} u^{\lambda} + \epsilon^{\lambda \nu \alpha \beta} u_{\alpha} B^a_{\beta}.
	\end{eqnarray}
	By the use of the second law of thermodynamics we find a frame in which the quark is not exerted by the drag force. We will see that this frame is exactly the Lab frame that we found by boosting the LL frame.

	Let us first consider the constitutive relations. Up to first order in derivatives and in a general hydrodynamic frame, which we call it X, we may write 
	  \begin{eqnarray}\label{T_no_darg}
	T^{\mu \nu} &=& w\, u^{\mu} u^{\nu} + p\, \eta^{\mu \nu} + \xi_{TB}^a \left(B^{a\mu} u^{\nu} + B^{a\nu} u^{\mu}\right) + \xi_{T\omega} \left(\omega^{\mu} u^{\nu} + \omega^{\nu} u^{\mu}\right),\\
	 J^{a \mu } &=& n^a u^{\mu}+\xi_{JB}^{ab}\,\, B^{b\mu} + \xi_{J\omega}^a\,\, \omega^{\mu},\\
	S^{\mu} &=&s\, u^{\mu}+ \xi_{sB}^a \,\,B^{a\mu} + \xi_{s\omega}\,\, \omega^{\mu}.
	\end{eqnarray}
	We have assumed that no dissipative effect is present in the fluid except for the  drag force on the quark.
	In  order to apply the second law of thermodynamics, we have to  compute the divergence of the entropy current. 
	To this end we first use the ideal fluid (zeroth order) equations of motion and obtain
	\begin{eqnarray}
	\partial_{\mu} B^{\mu}_a& =& - 2\omega \cdot E_a + \frac{B^{a\mu}}{w} \left( n^c E^c_{\mu } - \partial_{\mu}p + \zeta^c B^c_{\mu} +\frac{}{} \kappa \omega_{\mu} \right),\nonumber\\
	\partial_{\mu} \omega^{\mu} &=& \frac{2\omega^{\mu}}{w} \left( n^c E^c_{\mu} - \partial_{\mu}p +\frac{}{}\zeta^c B^c_{\mu} +\kappa  \omega_{\mu} \right),
	\end{eqnarray}
	with
	\begin{equation}
\zeta^c= \mathcal{D}\xi_{TB}^c + 2\theta \,\xi_{TB}^c,\,\,\,\,\,\,\,\,\,\,
\kappa=\mathcal{D}\xi_{T\omega} + 2\theta\, \xi_{T\omega}
	\end{equation}
	 Then by performing some lengthy computations we reach
	\begin{equation}\label{div_S}
	\begin{split}
	\partial_{\mu}S^{\mu} =\,u\cdot\mathcal{F}\,\,&+\,\, \left(  \partial_{\mu}\xi_{sB}^b +\frac{\mu^a}{T} \partial_{\mu}\xi_{JB}^{ab} -\frac{1}{T} \partial_{\mu}\xi_{TB}^b \right)B^{b\mu} + \left(\frac{1}{T}\xi_{JB}^{cb} - C^{abc}\frac{\mu^a}{T} \right) B^{b\mu}E^c_{\mu} \\
	&+ \,\frac{1}{w}\left(\xi_{sB}^b + \frac{\mu^a}{T } \xi_{JB}^{ab}  -\frac{2}{T}\xi_{TB}^b\right) \big(n^c B^{b\mu} E^c_{\mu} -B^{b\mu} \partial_{\mu}p + B^{b\mu} B^c_{\mu} \zeta^c + B^{b\mu} \omega_{\mu} \kappa\big)\\
	&+ \,\omega^{\mu} \left(  \partial_{\mu}\xi_{s\omega} + \frac{\mu^a}{T} \partial_{\mu}\xi_{J\omega}^a  - \frac{1}{T}\partial_{\mu}\xi_{T\omega}\right) + \left(\frac{1}{T}\xi_{J\omega}^b -2\frac{}{} \big( \,\xi_{sB}^b + \frac{\mu_a}{T} \xi_{JB}^{ab}  -\frac{1}{T}\xi_{TB}^b\big)\right) \omega^{\mu} E^b_{\mu}\\
	&+ \,\frac{1}{w}\left(\,\xi_{s\omega} + \frac{2\mu^a}{T} \xi_{J\omega}^a  -\frac{3}{T}\xi_{T\omega}\right) \big(n^c \omega^{\mu} E^c_{\mu} -\omega^{\mu} \partial_{\mu}p + \omega^{\mu} B^c_{\mu} \zeta^c + \omega^{\mu} \omega_{\mu} \kappa\big).
	\end{split}
	\end{equation}
	where we used the following definitions:
	  \begin{equation}
	\mathcal{D} \equiv u^{\nu} \partial_{\nu},~~~\theta \equiv \partial_{\mu} u^{\mu}.
	\end{equation}
	In the absence of $u\cdot\mathcal{F}$ term, there are nine independent structures in \eqref{div_S} which may have either positive or negative sign.  So in order to the divergence of entropy current be non-negative, one demands that these nine contributions have to vanish identically. However, these nine equations are not fully independent. It turns out that three of them  are combinations of the other six. So one finds six independent constraints as the following
\begin{eqnarray}\label{95}
\xi_{JB}^{ab} - C^{abc}\mu^c &=& 0,\\\label{96}
 T\xi_{sB}^b + \mu^a \xi_{JB}^{ab}  -2\xi_{TB}^b &=&0,\\\label{97}
2 (T \xi_{sB}^b + \mu^a \xi_{JB}^{ab} -\xi_{TB}^b) &=&\xi_{J\omega}^b, \\\label{98}
 2T\xi_{s\omega} + 2\mu^a \xi_{J\omega}^a  -3\xi_{T\omega}&=&0,\\\label{99}
T \mathcal{D} \xi_{sB}^b + \mu^a \mathcal{D}\xi_{JB}^{ab}  - \mathcal{D}\xi_{TB}^b &=& 0,\\\label{100}
 T \mathcal{D} \xi_{s\omega} + \mu^a \mathcal{D}\xi_{J\omega}^a  - \mathcal{D}\xi_{T\omega} &=& 0.
\end{eqnarray} 
Among these, the first four equations are algebraic while the last two ones are differential equations.  From equations \eqref{95} and \eqref{96} we have
\begin{eqnarray}\label{101}
	\xi_{JB}^{ab} = C^{abc} \mu^c,\,\,\,\,\,\,\,\,\,\,\,\,\,\,\xi_{TB}^{b}=\frac{C^{abc} \mu^a \mu^c}{2} + \frac{T \xi_{sB}^b}{2}.
\end{eqnarray}
Now we insert these expressions into the differential equation \eqref{99}. Recalling that the coefficients $C^{abc}$ are totally symmetric under permutation of indices, we find
\begin{eqnarray}
	T \mathcal{D} \xi_{sB}^b + \mu^a \mathcal{D}\xi_{JB}^{ab}  - \mathcal{D}\xi_{TB}^b = T \mathcal{D}\xi_{sB}^b - \xi_{sB}^b \mathcal{D}T=0.
\end{eqnarray}
This equation can be solved exactly for $\xi_{sB}^b$
\begin{eqnarray}\label{103}
T \mathcal{D}\xi_{sB}^b - \xi_{sB}^b \mathcal{D}T = T^2 \mathcal{D}\left(\frac{\xi_{sB}^b}{T}\right)=0 \,\,\,\,\rightarrow\,\,\,\,\,\,
 \xi_{sB}^b = \beta^b T.
\end{eqnarray}
with $\beta^a$ is a constant coefficient.
Then by substituting equations \eqref{101} and \eqref{103} in \eqref{97} we obtain
\begin{eqnarray}\label{104}
	\xi_{J\omega}^b= \beta^b T^2 + C^{abc} \mu^a \mu^c.
\end{eqnarray}
This together with \eqref{98} leads to 
\begin{equation}\label{105}
	\xi_{T\omega} = \frac{2}{3} C^{abc} \mu^a \mu^b \mu^c + \frac{2}{3} \beta^a \mu^a T^2 + \frac{2\,T \xi_{s\omega}}{3}.
\end{equation}
By substituting expressions \eqref{104} and \eqref{105} into the differential equation \eqref{100} we reach 
\begin{eqnarray}
	 T \mathcal{D} \xi_{s\omega} - 2 \xi_{s\omega} \mathcal{D} T + 2 \beta^a \mu^a T \mathcal{D} T - 2 \beta^a T^2 \mathcal{D} \mu^a =0,
\end{eqnarray}
which gives
\begin{equation}
	 T^3 \mathcal{D} \left(\frac{\xi_{s\omega}}{T^2} - \frac{2 \mu^a \beta^a}{T}\right) =0\,\,\,\,\,\,\rightarrow\,\,\,\,\,\,
	\xi_{s\omega} = \gamma T^2 + 2 \beta^a \mu^a T.
\end{equation}
with $\gamma$ a constant coefficient. 
In what follows we write the final form of  these  chiral transport coefficients in X frame
 \begin{eqnarray}\label{energy-component}
 \xi_{TB}^{a}& =& \frac{1}{2}\big(\beta^a T^2 + C^{abc} \mu^b \mu^c\big),\,\,\,\,\,\,\,\,\,\,\,\,\,\,\,\xi_{T\omega} = \frac{2C^{abc}}{3}\mu^a \mu^b \mu^c + 2\beta^a \mu^a T^2 + \frac{2\gamma}{3}T^3\label{sol_11},\\\label{j-component}
\hspace{-1cm} \xi_{JB}^{ab} &=& C^{abc}\mu^{c},\,\,\,\,\,\,\,\,\,\,\,\,\,\,\,\,\,\,\,\,\,\,\,\,\,\,\,\,\,\,\,\,\,\,\,\,\,\,\,\,\,\,\,\,\,\,\,\,\,\,\xi_{J\omega}^{a}= \beta^a T^2 + C^{abc} \mu^b \mu^c\label{sol_22},\\ \label{entropy-component}
~~~\xi_{sB}^{a} &=& \beta^a T, \,\,\,\,\,\,\,\,\,\,\,\,\,\,\,\,\,\,\,\,\,\,\,\,\,\,\,\,\,\,\,\,\,\,\,\,\,\,\,\,\,\,\,\,\,\,\,\,\,\,\,\,\,\,\,\,\,\,\xi_{s\omega} = 2\beta^{a} \mu^a T + \gamma T^2. ~~~\label{sol_33}
\end{eqnarray}
where as mentioned above, $\beta^a$ and $\gamma$ are constant. Substituting these solutions into \eqref{div_S}, we are left with 
\begin{equation}\label{div_S_uF}
\partial_{\mu}S^{\mu} =\,u\cdot\mathcal{F}.
\end{equation}

	The general structure of $\mathcal{F}^{\mu}$ for a quark moving in a neutral fluid has been found in   \cite{Abbasi:2012qz,Abbasi:2013mwa},  up to first order in derivative expansion. It turns out that in the absence of dissipation and in a non-chiral fluid, a quark with four velocity $\tilde{u}^{\mu}$ is dragged by the fluid with
\begin{equation}\label{zero_drag}
\mathcal{F}_{(0)}^{\mu}=\alpha_1 \left((u.\tilde{u})\,\tilde{u}^{\mu}+\frac{}{}u^{\mu}\right),
\end{equation}	
	where $\alpha_1$ is a non-negative coefficient function including the information about the interaction between quark and the fluid as well as the thermodynamics of the fluid; so it may depend on $S_1=u \cdot \tilde{u}$, $T$ and etc. In a charged fluid, however, the coefficient function might depend on the chemical $\mu^a$ potential as well.   In a slowly rotating chiral fluid coupled to a homogeneous weak magnetic field, the above drag force gets correction from both magnetic field and vosticity. Considering the set of all zero and first order derivative independent data  given in Table \ref{table_two}, we may formally write the most general form of the covariant chiral drag force  up to first order in derivative expansion in the X frame as
	\begin{equation}\label{general_F}
\begin{split}
\mathcal{F}^{\mu}&=\mathcal{F}_{(0)}^{\mu}+\epsilon_f\,\mathcal{F}_{(1)}^{\mu}
\\
&=\left(1+\epsilon_f\,\frac{}{}\big(\alpha^{a}_{s2}\,S^a_{2}+\alpha_{s3}\,S_{3}\big)\right)\mathcal{F}_{(0)}^{\mu}\,+\,\epsilon_f\,\left(\alpha^{a}_{v2}\,V^{a \mu}_{2}+\frac{}{}\alpha_{v3}\,V^{\mu}_{3}\right).
\end{split}
	\end{equation}
	Here $\mathcal{F}_{(0)}^{\mu}$ is nothing but the drag force in the ideal fluid, i.e. \eqref{zero_drag}, which may be also written as 
	\begin{equation}\label{zero_drag'}
	\mathcal{F}_{(0)}^{\mu}=\alpha_1\,V_{1}^{\mu}.
	\end{equation}	
		\begin{table}[!htb]
		\label{table_two}
		\begin{center}
			\begin{tabular}{|c|c|c|}
				\hline
				\hline
				$SO(3)$  &  zero order   & first order   \\
			classification &    independent data &   independent data \\
				\hline
				\hline
				&  & \\ 
				
			Scalar   & $S_1=\,	u\cdot \tilde{u}$&  $S_2^a=\,	 \tilde{u}\cdot B^a$   \\
				
				&   &$S_3=\,	 \tilde{u}\cdot \omega$  \\
				& & \\
				\hline
				\hline
				  &  &  \\
				
				Vector     & $V_1^{\mu}=\, u^{\mu}+S_1\,\tilde{u}^{\mu}$ &  \,\,\,\,\,\,\,$V_2^{a\mu}=\, B^{a\mu}+S_2^a\,\tilde{u}^{\mu}$  \,\,\,\,\,  \\
				&   & \,\,\,\,\,$V_3^{\mu}=\, 
				\omega^{\mu}+S_3\,\tilde{u}^{\mu}$\,\,\,\,\, \\
			& & \\
				\hline
				\hline
			\end{tabular}
		\end{center}
		\caption{One derivative data contributing to chiral drag force}
		\label{table_two}
	\end{table}
	It is clear that once the five drag coefficient  functions, namely $\alpha$, $\alpha_{s2}^a$, $\alpha_{s3}$, $\alpha_{v2}^a$ and $\alpha_{v3}$, are determined, the drag force would be fully specified.  What may constrain the structure of these coefficient functions is the second law of thermodynamics which states that the divergence of entropy current associated with any physical hydrodynamical flow must be non-negative.
	Let us so simplify \eqref{div_S_uF} by use of \eqref{general_F}. It turns out that
	\begin{equation}
\partial_{\mu}S^{\mu} =\,u\cdot\mathcal{F}=(1+\alpha_{s2}^a\, S_2^a+\alpha_{s3}\, S_3)(S_1^2-1)\,\alpha_1+\,\alpha_{v2}^a\, S_1S_2^a+\,\alpha_{v3}\, S_1S_3.
	\end{equation}
	We already know that the second law constrains the coefficient of drag in the ideal fluid  as $\alpha_1\ge0$. However, it would not be possible to constrain the other four coefficients by the second law. This is  due to this fact that the $S_2^a$ and $S_3$ may have either sign,  depending on the  profile of magnetic field and vorticity. So we conclude that in order to hold solutions \eqref{sol_11}, \eqref{sol_22}, \eqref{sol_33} and simultaneously     satisfy $\partial_{\mu}S^{\mu}\ge0$,  the four unknown coefficients of the chiral drag force must vanish in the X frame:
	\begin{equation}
\alpha_{s2}^a=\,\alpha_{s3}=\,\alpha_{v2}^a=\,\alpha_{v3}=\,0.
	\end{equation}  
The chiral drag force then turns out to take the  following form in the X frame:
\begin{equation}\label{drag_X}
\mathcal{F}_{\text{X}}^{\mu}=\alpha_1 \left((u.\tilde{u})\,\tilde{u}^{\mu}+\frac{}{}u^{\mu}\right),\,\,\,\,\,\,\alpha_1\ge0.
\end{equation}	
	Now let us investigate  how a quark at rest, $\tilde{u}^{\mu}=(1,\boldsymbol{0})$, in the rest frame of the fluid, $u^{\mu}=(1,\boldsymbol{0})$, in X frame, is dragged. It is simple to see that in this case \eqref{drag_X} simplifies to
	\begin{equation}
\mathcal{F}_{\text{X}}^{\mu}=\,0,
	\end{equation}
	which means that \textbf{X frame is nothing but the no-drag frame}. 
	
When the fluid is constituted of two U(1) flavors, i.e. one axial $A$ together  with one vector $V$, then $a,b,c\in\{A,V\}$  and we will have
\begin{eqnarray}\label{inice}
C^{VVV}= C^{AAV}=0,\,\,\,\,\,\,\,C^{AAA}= C^{VVA}=\mathcal{C},
\end{eqnarray}
and also $\beta^A=0$ and $\beta^V=\mathcal{D}$.
The structure of constitutive relations in this case may be written as
	\begin{eqnarray}
	\label{Tmunu_Lab}
	T^{\mu \nu} &= & w\, u^{\mu} u^{\nu} + p\, g^{\mu \nu}+\sigma^{\mathcal
		\epsilon}_{\mathcal{B}}(u^{\mu}B^{\nu}+u^{\nu}B^{\mu})+\sigma^{\mathcal
		\epsilon}_{\mathcal{\omega}}(u^{\mu}\omega^{\nu}+u^{\nu}\omega^{\mu}),\\
	\label{jmu_Lab}
	J_V^{\mu} & = & n_V\, u^{\mu} +\sigma_{V\mathcal{B}}\, B^{\mu}+\sigma_{V\mathcal{\omega}}\,\omega^{\mu},\\\label{jmu5_Lab}
	J_A^{\mu} & = & n_A\,u^{\mu} +\sigma_{A\mathcal{B}}\, B^{\mu}+\sigma_{A\mathcal{\omega}}\omega^{\mu}.
	\end{eqnarray}
	The corresponding transport coefficients are given by \cite{Gao:2012ix,Landsteiner:2012kd,Landsteiner:2016led}
	\begin{eqnarray}\label{trans_lab1}
	\sigma^{\mathcal
		\epsilon}_{\mathcal{B}}&=&\frac{\bar{\boldsymbol{\chi}}_{R}-\bar{\boldsymbol{\chi}}_{L}}{4 }=\frac{\mu_R^2 - \mu_L^2}{8\pi^2},\\\label{trans_lab_2}
	\sigma^{\mathcal
		\epsilon}_{\mathcal{\omega}}&=&n_R-n_L,\\
	\sigma_{V\mathcal{B}}&=&\left(\xi_{VB}+\,n_V\,\frac{\bar{\boldsymbol{\chi}}_{R}-\bar{\boldsymbol{\chi}}_{L}}{4 w}\right)=\,\frac{\mu_R-\mu_L}{4\pi^2},\\
	\sigma_{V\mathcal{\omega}}&=&\left(\xi_{VB}+\,n_V\,\frac{n_R-n_L}{ w}\right)=\,\frac{\mu_R^2-\mu_L^2}{4\pi^2},\\
	\sigma_{A\mathcal{B}}&=&\left(\xi_{AB}+\,n_A\,\frac{\bar{\boldsymbol{\chi}}_{R}-\bar{\boldsymbol{\chi}}_{L}}{4 w}\right)=\,\frac{\mu_R+\mu_L}{4\pi^2}\label{trans_lab2},\\
	\sigma_{A\mathcal{\omega}}&=&\left(\xi_{AB}+\,n_A\,\frac{n_R-n_L}{w}\right)=\,\frac{T^2}{6}+\,\frac{\mu_R^2+\mu_L^2}{4\pi^2}.
	\end{eqnarray}
	In the following we show that these set of chiral transport coefficients of the Lab frame, are a special case of coefficients \eqref{sol_11}, \eqref{sol_22} and \eqref{sol_33} of the no-drag frame. First by the use of of \eqref{inice}, we rewrite the mentioned transport coefficients as the following 
	\begin{eqnarray}\label{}\label{}
	\xi_{TB}^{V}& =& \frac{1}{2}\big(\beta^V T^2 + \mathcal{C} \mu_V \mu_A\big),\,\,\,\,\,\,\,\,\,\,\,\,\,\,\,\xi_{T\omega} \,=\, \frac{2\mathcal{C}}{3}\big(\mu_V^2+3\mu_A^2\big) + 2\big(\beta^V \mu^V+\beta^A \mu^A\big) T^2 + \frac{2\gamma}{3}T^3\label{},\\
\xi_{JB}^{AV} &=&\mathcal{C}\mu_{V},\,\,\,\,\,\,\,\,\,\,\,\,\,\,\,\,\,\,\,\,\,\,\,\,\,\,\,\,\,\,\,\,\,\,\,\,\,\,\,\,\,\,\,\,\,\,\,\,\,\,\,\,\,\,\,\xi_{JB}^{VV}=\,\mathcal{C}\mu_{A},\\ 
\xi_{J\omega}^{A}&=&\beta^A T^2 +\mathcal{C} \big(\mu_A^2+\mu_V^2\big),\label{}\,\,\,\,\,\,\,\,\,\,\,\,\,\,\xi_{J\omega}^{V}\,=\,\beta^V T^2 + \mathcal{C} \big(\mu_V^2+3\mu_A^2\big),\\
\xi_{sB}^{V}& =& \beta^V T, \,\,\,\,\,\,\,\,\,\,\,\,\,\,\,\,\,\,\,\,\,\,\,\,\,\,\,\,\,\,\,\,\,\,\,\,\,\,\,\,\,\,\,\,\,\,\,\,\,\,\,\,\,\,\,\,\,\,\xi_{s\omega}\, =\, 2\big(\beta^{V} \mu_V+\beta^{A} \mu_A\big) T + \gamma\, T^2. 
\end{eqnarray}
	Equating each of the above expressions with its counterpart among equations \eqref{trans_lab1} to \eqref{trans_lab2}, we obtain 
	\begin{equation}\label{const_coef}
		\gamma=0,\,\,\,\,\,\,\,\,\,\,\,\,\,\beta^V=0,\,\,\,\,\,\,\,\,\,\beta^A=\frac{1}{6},\,\,\,\,\,\,\,\,\,\,\,\,\mathcal{C}=\,\frac{1}{2\pi^2}.
	\end{equation}
	That we could find a set of constant coefficients \eqref{const_coef} which equates the transport coefficients in the Lab frame with those in no-drag frame shows that the Lab frame is basically a no-drag frame.
	
	Let us review what we found in this subsection The novel result was 
	finding the anomalous transport coefficients  in the no-drag frame for a system of non-Abelian fermions. As it was expected, we also showed that the system under our study in the first part of the paper, namely the system of right- and left-handed fermions, in the hydro limit is identified in a no-drag frame.

	\subsection{Reproducing the Results of Neiman-Oz by Boost}
	As the first check for the idea of boost, in section \ref{sec_5},  we showed that how the velocity of hydro modes in LLframe may be found from those in the Lab frame.  In this subsection we check the idea of boost in another way. We focus on the system of non-Abelian chiral fermions. The structure of the hydro constitutive relations for this system was completely specified in no-drag frame in previous subsection.  After finding the boost velocity, we make a boost from the rest frame of the fluid in no-drag frame to the rest frame of the fluid in LL frame. We then find the constitutive relations in the latter frame and compare them with the results of \cite{Neiman:2010zi}.

	Motivated by observation of the difference between \eqref{T_0i_CKT}
	and \eqref{T_0i_LL}, we first compute the energy flow in the rest frame of the fluid in no-drag frame. Let us take $u^{\mu}=(1,\boldsymbol{0})$ in \eqref{T_no_darg}. We obtain
\begin{eqnarray}\label{T_rest_no_drag}
T^{00}&=&\,\epsilon,\,\,\,\,\,\,\,\,\,\,T^{0i}=\,\xi_{TB}^a\,B^{ai}  + \xi_{T\omega}\, \Omega^{i},\,\,\,\,\,\,\,\,\,\,T^{ij}=p\delta^{ij}\\ \label{J_rest_no_drag}
J^{0}&=&n,\,\,\,\,\,\,\,\,\,\,J^{ai}=\,\xi_{JB}^{ab}\,\, B^{bi} + \xi_{J\omega}^a\,\, \Omega^{i}
\end{eqnarray}
	According to the discussion of section \ref{sec_5}, the velocity of the boost between no-drag frame and LL frame will then be
	\begin{equation}\label{boost_non_Abelian}
\beta_i=\,\frac{1}{w}T^{0i}=\,\frac{\xi_{TB}^a}{w}\,B^{ai}  + \frac{\xi_{T\omega}}{w}\, \Omega^{i}.
	\end{equation}
Now by applying the above boost to \eqref{T_rest_no_drag} and \eqref{J_rest_no_drag}, we obtain (see \ref{ppendix_boost} for details)
\begin{eqnarray}\label{T_rest_LL}
T^{00}&=&\,\epsilon,\,\,\,\,\,\,\,\,\,\,T^{0i}=\,0\, ,\,\,\,\,\,\,\,\,\,\,\,\,\,\,\,\,\,\,\,\,T^{ij}=p\delta^{ij}\\ \label{J_rest_LL}
J^{0}&=&n,\,\,\,\,\,\,\,\,\,\,J^{ai}=\, \left(\xi_{JB}^{ab} -\frac{ \xi_{TB}^b n^a}{w}\right) B^{ai}  + \left(\xi_{J\omega}^a - \frac{\xi_{T\omega} n^a}{w}\right)\, \Omega^{i}.
	\end{eqnarray}
So in a general Lorentz frame we can write
\begin{eqnarray}
T^{\mu\nu}_{LL}&=&\,(\epsilon+p)u^{\mu}u^{\nu}+p \eta^{\mu\nu}\\	
J^{a\mu}_{LL}&=& n^{a} u^{\mu} + \left(\xi_{JB}^{ab} -\frac{ \xi_{TB}^b n^a}{w}\right) B^{b\mu} +  \left(\xi_{J\omega}^a - \frac{\xi_{T\omega} n^a}{w}\right) \omega^\mu.
\end{eqnarray}
	By using \eqref{energy-component}, \eqref{j-component} and \eqref{entropy-component} we find the coefficients of CME and CVE in the LL frame as the following
\begin{eqnarray}\label{trans-Lan}
&&\xi^{ab}_{B\,LL}=\xi_{JB}^{ab} -\frac{ \xi_{TB}^b n^a}{w} = C^{abc} \mu^c -\frac{n^a} {w} \left( \frac{\beta^b T^2}{2} + \frac{C^{bcd} \mu^c \mu^d}{2}\right),\nonumber\\
&&\xi^{a}_{\omega LL}=\xi_{J\omega}^a - \frac{\xi_{T\omega} n^a}{w}= \beta^a T^2 + C^{abc} \mu^b \mu^c -\frac{2n^a}{w} \left(\frac{C^{abc}}{3}\mu^a \mu^b \mu^c + \beta^a \mu^a T^2 + \frac{\gamma}{3}T^3\right).
\end{eqnarray}
These expressions are exactly the constitutive relations in LL frame found in \cite{Neiman:2010zi} for a system of
non-Abelian fermions.

\section{The General Form of Chiral Drag Force in the Landau-Lifshitz Frame: Comparison with the Holography Result }
\label{7}
In section \ref{sec_5}, we obtained the first evidence for testing this idea that the hydrodynamic frames may be transformed to each other by boost transformations.
We used these transformations to derive the hydrodynamic constitutive relations in the hydro limit of chiral kinetic theory from those of LL frame. Then in section \ref{sec_6}, we showed that the fluid frame associated with the hydro limit of chiral kinetic theory, namely the Lab frame, is actually a no-drag frame.  This observation motivates to examine the idea of boost by computing the chiral drag force in the LL frame and compare it with the well-known result obtained from the Fluid/Gravity duality in the same frame.

\subsection{Chiral Drag Force in the Landau-Lifshitz Frame}
 Our strategy to find the chiral drag force in the LL frame is as the following. 
 We first evaluate the general drag force  \eqref{general_F} in the rest frame of the fluid, $u^{\mu}=(1,\boldsymbol{0})$, for a quark moving with four velocity $\tilde{u}^{\mu}_{LL}=\tilde{\gamma}'(1,\tilde{v}'_x,\tilde{v}'_y,\tilde{v}'_z)$. We call this force as $\mathcal{F}^{\mu}_{RF,LL}$ \footnote{Note that indeed, this is the minus of the drag force.}. Considering \eqref{general_boost}, we then make a $-\boldsymbol{\beta}$ boost to
 go  from the    LL frame to the no-drag frame. Doing so, the four velocity of the quark is transformed to $\tilde{u}^{\mu}_{no-drag}=\tilde{\gamma}(1,\tilde{v}_{x},\tilde{v}_{y}, \tilde{v}_{z})$. The resultant drag force would be $\mathcal{F}^{\mu}_{RF,no-drag}$ and when evaluating for a quark at rest in this frame, has to vanish.
 This requirement gives the complete form of the chiral drag force in the rest frame of the fluid in the LL frame, consistent with the result of Fluid/Gravity duality.

 For the sake of simplicity we first focus on the non-rotating case and just couple the fluid to a magnetic field. The generalization to the case of rotating fluid would be straightforward.  We get the magnetic field in the $z-$ direction and so $B^{a\mu}=(0,0,0,\text{B}^a)$. By exploiting the rotational symmetry we can get the velocity of the quark as $\tilde{u}^{\mu}_{LL}=\tilde{\gamma}'(1,\tilde{v}'_x, 0,\tilde{v}'_z)$. The chiral drag force $\mathcal{F}^{\mu}_{RF,LL}$ then turns out to be 
 \begin{equation}\label{general_drag_RF}
 \mathcal{F}_{RF,LL}^{\mu}	=\,\begin{pmatrix}
 \alpha_{1}(-\tilde{\gamma}')\left(1-\tilde{\gamma}'^2+\frac{}{} \alpha_{s2}^a(-\tilde{\gamma}')\,(1-\tilde{\gamma}'^2)\tilde{\gamma}' \tilde{v}'_{z}\frac{}{}\text{B}^a\right )+\alpha_{v2}^a(-\tilde{\gamma}')\,\tilde{\gamma}'^2 \tilde{v}'_{z}\frac{}{}\text{B}^a\\
 \alpha_{1}(-\tilde{\gamma}')\left(-\tilde{\gamma}'^2\tilde{v}'_x-\frac{}{} \alpha^a_{s2}(-\tilde{\gamma}')\,\tilde{\gamma}'^3 \tilde{v}'_{x}\tilde{v}'_{z}\frac{}{}\text{B}^a\right )+\alpha_{v2}^a(-\tilde{\gamma}')\,\tilde{\gamma}'^2 \tilde{v}'_{x}\tilde{v}'_{z}\frac{}{}\text{B}^a \\
 0\\
 \alpha_{1}(-\tilde{\gamma}')\left(-\tilde{\gamma}'^2\tilde{v}'_z-\frac{}{} \alpha^a_{s2}(-\tilde{\gamma}')\,\tilde{\gamma}'^3 \tilde{v}'^2_{z}\frac{}{}\text{B}^a\right )+\alpha_{v2}^a(-\tilde{\gamma}')\,(1+\tilde{\gamma}'^2 \tilde{v}'^2_{z})\frac{}{}\text{B}^a\\
 \end{pmatrix}+O(\text{B}^2).
 \end{equation}
 In this expression we have explicitly shown the dependence of $\alpha_1$, $\alpha^a_{s2}$ and $\alpha^a_{v2}$
 on the $S_1=u\cdot\tilde{u}=-\tilde{\gamma}$. Note that, however, it is just for our later requirements and does not mean that these coefficient functions are functions of only $S_1$. As mentioned earlier, in general, they might depend on thermodynamic quantities like $T$ as well.
 
 Now, making a boost  $\beta^{\mu}=(1,0,0,-\beta)$ we find the drag force in the no-drag frame. From \eqref{general_boost} we know that $\beta \sim O(\partial)$ and so the boost velocity is normalized to $-1$, up to first order in derivatives, i.e. $\beta^{\mu}\beta_{\mu}=-1+O(\partial^2)$. Then for this special choice of rotational frame \eqref{general_drag_RF}  reads 
 \begin{equation}\label{drag_RF_no_drag_0}
 \mathcal{F}_{RF,no-drag}^{\mu}	=\,\begin{pmatrix}
 \alpha_{1}(-\tilde{\gamma}')\left(1-\tilde{\gamma}'^2+\frac{}{} \alpha^a_{s2}(-\tilde{\gamma}')\,(1-\tilde{\gamma}'^2)\tilde{\gamma}' \tilde{v}'_{z}\frac{}{}\text{B}^a\right )+\alpha_{v2}^a(-\tilde{\gamma}')\,\tilde{\gamma}'^2 \tilde{v}'_{z}\frac{}{}\text{B}^a- \alpha_1(-\tilde{\gamma}')\tilde{\gamma}'^2\tilde{v}_z'\,\beta\\
 \alpha_{1}(-\tilde{\gamma}')\left(-\tilde{\gamma}'^2\tilde{v}'_x-\frac{}{} \alpha^a_{s2}(-\tilde{\gamma}')\,\tilde{\gamma}'^3 \tilde{v}'_{x}\tilde{v}'_{z}\frac{}{}\text{B}^a\right )+\alpha_{v2}^a(-\tilde{\gamma}')\,\tilde{\gamma}'^2 \tilde{v}'_{x}\tilde{v}'_{z}\frac{}{}\text{B}^a \\
 0\\
 \alpha_{1}(-\tilde{\gamma}')\left(-\tilde{\gamma}'^2\tilde{v}'_z-\frac{}{} \alpha^a_{s2}(-\tilde{\gamma}')\,\tilde{\gamma}'^3 \tilde{v}'^2_{z}\frac{}{}\text{B}^a\right )+\alpha_{v2}^a(-\tilde{\gamma}')\,(1+\tilde{\gamma}'^2 \tilde{v}'^2_{z})\frac{}{}\text{B}^a+\alpha_1(-\tilde{\gamma}')(1-\tilde{\gamma}'^2)\,\beta\\
 \end{pmatrix}+O(\text{B}^2).
 \end{equation}
 In order to write the above force in terms of the quark velocity in the no-drag frame, we boost the quark velocity itself as the following:
 \begin{eqnarray}
 \tilde{v}_x'&=&\tilde{v}_x+\tilde{v}_x\tilde{v}_z\beta+\,O(\beta^2),\\
 \tilde{v}_y'&=&\tilde{v}_y+\,O(\beta^2),\\
 \tilde{v}_z'&=&\tilde{v}_z-(1-\tilde{v}_z^2)\beta+\,O(\beta^2),\\
 \tilde{\gamma}'&=&\tilde{\gamma}(1-\tilde{v}_z\,\beta)+\,O(\beta^2).
 \end{eqnarray}
 So \eqref{drag_RF_no_drag_0} may be rewritten as it follows:
 \begin{equation}\label{drag_RF_no_drag}
 \begin{split}
 &\mathcal{F}_{RF,no-drag}^{\mu}\\
 &	=\begin{pmatrix}
 \alpha_{1}(-\tilde{\gamma})+(1-\tilde{\gamma}^2)\left(\alpha_{1}(-\tilde{\gamma}) \alpha^a_{s2}(-\tilde{\gamma})\frac{}{}(1-\tilde{\gamma}^2)\tilde{\gamma}+\alpha_{v2}^a(-\tilde{\gamma})\,\tilde{\gamma}^2\right) \tilde{v}_{z}\frac{}{}\text{B}^a- \left(2\alpha_1(-\tilde{\gamma}) \tilde{\gamma}+\alpha'_1(-\tilde{\gamma})(1-\tilde{\gamma}^2)\right)\tilde{\gamma}\tilde{v}_z\,\beta\\
 -\alpha_{1}(-\tilde{\gamma})\tilde{\gamma}^2\tilde{v}_x+\left(- \alpha_{1}(-\tilde{\gamma}) \alpha^a_{s2}(-\tilde{\gamma})\tilde
 \gamma^3+\frac{}{}\alpha^a_{v2}(-\tilde{\gamma})\tilde{\gamma}^2\right)\,\tilde{v}_x\tilde{v}_z \text{B}^a+\left(\alpha_1(-\tilde{\gamma})-\frac{}{}\alpha'_1(-\tilde{\gamma})\tilde{\gamma}\,\right)\tilde{\gamma}^2 \tilde{v}_{x}\tilde{v}_{z}\frac{}{}\beta  \\
 0\\
 -\alpha_{1}(-\tilde{\gamma})\tilde{\gamma}^2\tilde{v}_z+\,\left(\alpha_{v2}^a(-\tilde{\gamma})+\left(\alpha_{v2}^a(-\tilde{\gamma})\tilde{\gamma}^2- \alpha_{1}(-\tilde{\gamma}) \alpha^a_{s2}(-\tilde{\gamma})\tilde
 \gamma^3\right)\tilde{v}_z^2 \right)\text{B}^a\,+\big(\left(\alpha_1(-\tilde{\gamma})-\alpha'_1(-\tilde{\gamma})\tilde{\gamma}\,\right)\tilde{\gamma}^2 \tilde{v}_{z}^2-\alpha_1(-\tilde{\gamma})\tilde{\gamma}^2\big)\beta\\
 \end{pmatrix}
 \end{split},
 \end{equation}
 where 
 \begin{equation}
 \alpha_1'(-\tilde{\gamma})=\,\left(\frac{\partial \alpha_1}{\partial S_1}\right)_{S_1=-\tilde{\gamma}}.
 \end{equation}
 For a quark at rest in the rest frame of the fluid, \eqref{drag_RF_no_drag} takes a simpler form which we call it $\mathcal{F}_{RF,no-drag}^{\mu,rest-q}$ in the following. Let us take $\tilde{v}_x=\tilde{v}_z=0$ and consequently $\tilde{\gamma}=1$.
 We then find 
  \begin{equation}
 \mathcal{F}_{RF,no-drag}^{\mu,rest-q}	=\,\begin{pmatrix}
  0 \\
0 \\
  0\\
  \alpha_{v2}^a(-1)\text{B}^a-\alpha_1(-1)\beta
  \end{pmatrix}.
  \end{equation}
  Since by definition, $ \mathcal{F}_{RF,no-drag}^{\mu,rest-q}$ has to vanish, we obtain the following constraint on the coefficient functions of drag force:
  \begin{equation}
\alpha_{v2}^a(-1)\text{B}^a=\,\beta\,\alpha_1(-1).
  \end{equation}
  Let us now focus on the special case which the system  contains the charges of the symmetry group $U(1)_A\times U(1)_V$. We get naturally $\text{B}^A=0$ and $\text{B}^V=\text{B}$.  We also drop the superscript of $\alpha_{s2}^a$ and $\alpha_{v2}^a$.
  If we added a global rotation we would also find a similar constraint on $\alpha_{v3}(-1)$. On the other hand   by use of  \eqref{general_boost}, \eqref{trans_lab1} and \eqref{trans_lab_2}, $\beta$ is given by
  \begin{equation}\label{boost_final}
  \boldsymbol{\beta}=\frac{\sigma^{\mathcal
  		\epsilon}_{\mathcal{B}}}{4 w}\,\textbf{B}+\,\frac{\sigma^{\mathcal
  		\epsilon}_{\mathcal{\omega}}}{w}\,\boldsymbol{\Omega}.
  \end{equation}
  So one concludes that
  \begin{eqnarray}\label{constraint_1}
  \alpha_{v2}(-1)&=&\,\frac{\sigma^{\mathcal
  		\epsilon}_{\mathcal{B}}}{4 w}\,\alpha_1(-1),\\\label{constraint_2}
  	 \alpha_{v3}(-1)&=&\,\frac{\sigma^{\mathcal
  			\epsilon}_{\mathcal{\omega}}}{ w}\,\alpha_1(-1).
  \end{eqnarray}
 Up to know and in general, what all we have found about the chiral drag force in the LL frame is the general form of the drag force \eqref{general_F} together with two constraints \eqref{constraint_1} and \eqref{constraint_2}. In the next subsection we show that these are sufficient to reproduce the AdS/CFT result on drag force exerted on a quark at rest in the rest frame of the fluid in LL frame.
 \subsection{Comparison with the AdS/CFT Result}
 Computing drag force exerted on a moving quark in a dynamical flow is the subject of series of papers \cite{Abbasi:2012qz,Abbasi:2013mwa,Lekaveckas:2013lha}. Using Fluid/Gravity duality, the effect of fluid gradients on the drag force was computed in these papers. It has been shown that when the fluid is neutral, the effect arises in terms of $\partial_{\mu}u^{\mu}$ and $\partial_{\mu}T$. For a charged fluid, however, terms with gradients of chemical potentials, like $\partial_{\mu}\frac{\mu}{T}$, appear as well. All these corrections are dissipative corrections at first order in derivatives.
 
 Such computations has been extended to the case of a chiral fluid. In 	\cite{Rajagopal:2015roa} by considering a globally rotating fluid in presence of a magnetic field, the chiral drag force is computed to first order in derivative expansion. The latter means that the magnetic field and vorticity of the fluid have been assumed such weak that one can regard them as one derivative objects in derivative expansion.  In the following, we will demonstrate that by use of the well-known results of AdS/CFT duality (and Fluid/Gravity duality) on $\alpha_1$, $\sigma^{\mathcal
 	\epsilon}_{\mathcal{B}}$ and $\sigma^{\mathcal
 	\epsilon}_{\mathcal{\omega}}$, one can successfully reproduce the chiral drag force exerted on a quark at rest in the rest frame of the fluid in the LL frame, computed in \cite{Rajagopal:2015roa}, with no use of the holographic computations done at that paper.

Let us compute the chiral drag force exerted on the quark in the rest frame of the LL frame. Considering \eqref{general_drag_RF}, with $\tilde{v}_x=\tilde{v}_z=0$ and also by adding a global vorticity $\omega$, one finds:
 \begin{equation}
 \mathcal{F}_{RF,LL}^{\mu,rest-q}	=\,\begin{pmatrix}
 0\\
 0 \\
 0\\
 \alpha_{v2}(-1)\,\frac{}{}\text{B}+\alpha_{v3}(-1)\,\frac{}{}\Omega\\
 \end{pmatrix}+O(\text{B}^2, \Omega^2).
 \end{equation}
 Now by using \eqref{constraint_1} and \eqref{constraint_2}, we may write the chiral drag 3-force covariantly as 
 \begin{equation}\label{drag_cov_us}
 \boldsymbol{f}=\,\alpha_1(-1)\left(\frac{\sigma^{\mathcal
 		\epsilon}_{\mathcal{B}}}{4 w}\, \textbf{B}+\frac{\sigma^{\mathcal
 		\epsilon}_{\mathcal{\omega}}}{ w}\,\boldsymbol{\Omega}\right).
 \end{equation}

 As mentioned below \eqref{zero_drag}, the $\alpha_1$ coefficient in general is a function of $S_1$ as well as thermodynamic quantities like $T$  and $\mu$.
 For quark moving in a hot plasma like quark gluon plasma, due to strong coupling, perturbative computations fail. Alternatively, by using the AdS/CFT and holographic pictures, some analytical results have been found about the drag force. In 	\cite{Gubser:2006bz,Herzog:2006gh} the drag force affected on a moving quark in a neutral static plasma has been computed via attaching the quark to the end point of a string trailing in the AdS bulk and studying the dynamics on the world-sheet. By making a boost and go to a general Lorentz frame, one may generally find the drag force exerted on a quark with velocity $\tilde{u}^{\mu}$ in a fluid with velocity $u^{\mu}$, namely the force given by  \eqref{zero_drag}. It turns out that 
 \begin{equation}
 \mathcal{F}_{(0)}^{\mu}=\,\frac{\sqrt{\lambda}}{2 \pi}\,\pi^2 T^2 \left((u.\tilde{u})\,\tilde{u}^{\mu}+\frac{}{}u^{\mu}\right)
 \end{equation}
 where $\lambda$ is the t'Hooft coupling. The author of 		\cite{Herzog:2006se} generalized such computations to the case in which a $U(1)$ global charge is in the system as well, with its corresponding chemical potential $\mu$.  It was shown that  in this case and to order $\mu^2/T^2$:
 \begin{equation}
 \mathcal{F}_{(0)}^{\mu}=\,\frac{\sqrt{\lambda}}{2 \pi}\,\pi^2 T^2\left(1+\frac{1+3 S_1}{6S_1} \frac{\mu^2}{(\pi T)^2}\right) \left((u.\tilde{u})\,\tilde{u}^{\mu}+\frac{}{}u^{\mu}\right).
 \end{equation}
 It is natural to more generalize this formula for  a plasma with  a number of $U(1)$ global currents. When considering right and left handed charges, one may replace $\mu^2$ with $\mu_R^2-\mu_L^2$ or in the basis of vector and axial currents with $4\mu_A\mu_V$. So, what AdS/CFT gives as $\alpha_1$ for our purposes in this paper, up to order $\mu^2/T^2$, is as the following:
 \begin{equation}\label{alpha_1}
 \alpha_1=\,\frac{\sqrt{\lambda}}{2 \pi}\,\pi^2 T^2\left(1+\frac{1+3 S_1}{6S_1} \frac{4\mu_A\mu_V}{(\pi T)^2}\right).
 \end{equation}
 On the other hand, $\sigma^{\mathcal
 	\epsilon}_{\mathcal{B}}$ and $\sigma^{\mathcal
 		\epsilon}_{\mathcal{\omega}}$ have been found in the AdS/CFT as well. Adding the Chern-Simons coupling terms for both $F^2$ and $R^2$ terms in the AdS bulk, the authors of  		\cite{Landsteiner:2012kd} have found these coefficients in a fluid with a single anomalous current as
 	\begin{eqnarray}
 	\sigma_{\mathcal{B}}^{\epsilon}&=&\,\frac{\pi}{2 G_5}\left(\kappa\,\frac{\mu^2 }{2 \pi^2}+\kappa_g\,4T^2\right)\\
 		\sigma_{\mathcal{\omega}}^{\epsilon}&=&\,\frac{\pi}{2 G_5}\left(\kappa\,\frac{\mu^3 }{3 \pi^2}+\kappa_g\,8 \mu T^2\right)
 	\end{eqnarray}
 where $\kappa$ and $\kappa_g=\kappa/24$ are proportional to the  coefficients of axial anomaly and gravitational anomaly on the boundary field theory, respectively. In the expressions above, $G_5$ is the Newton constant in 5-dimension and can be written in terms of the boundary theory thermodynamics 		\cite{CasalderreySolana:2011us}. Doing so in a system of both vector and axial currents, we obtain  
 \begin{eqnarray}\label{sigma_B_holog}
 \sigma_{\mathcal{B}}^{\epsilon}&=&\frac{w}{(\pi T)^2}\left(\frac{\kappa}{2 \pi^2}\,\frac{4\mu_A\mu_V }{T^2}+8\kappa_g\right)\\ \label{sigma_omega_holog}
 \sigma_{\mathcal{\omega}}^{\epsilon}&=&\frac{w}{(\pi T)^2}\left(\frac{2 \kappa}{3\pi^2}\,\frac{6 \mu_V^2+2\mu_A^2 }{T^2}+16\kappa_g\right)\mu_A.
 \end{eqnarray}
  Substituting \eqref{alpha_1}, \eqref{sigma_B_holog} and \eqref{sigma_omega_holog} into \eqref{drag_cov_us} and keeping terms up to second order on $\mu/T$, we obtain:
  \begin{equation}\label{last_result}
\begin{split}
\boldsymbol{f}=\,\frac{\kappa\sqrt{\lambda}}{2\pi^3}\left(\frac{4\mu_A\mu_V }{T^2}+16\pi^2\frac{\kappa_g}{\kappa}\right)\textbf{B}+\,\frac{\kappa\sqrt{\lambda}}{3\pi^3}\left(\frac{6 \mu_V^2+2\mu_A^2  }{T^2}\mu_A+24\pi^2\frac{\kappa_g}{\kappa}\,\mu_A\right)\boldsymbol{\Omega}.
\end{split}
  \end{equation}
 This is one last result in this paper. There are two points however which has to be noted. First, when $\kappa_g=0$, this formula is exactly coincides with the formula (5.1) derived in \cite{Rajagopal:2015roa}. The apparent difference in the multiplicative factor in second term is due to this fact that the definition of vorticity in \cite{Rajagopal:2015roa} is as   $\omega^{\mu}=\epsilon^{\mu\nu\alpha\beta}u_{\nu}\partial_{\alpha}u_{\beta}$ which is twice of our vorticity defined below \eqref{jmu5_LL}.
 Second, we have not only reproduced the result of \cite{Rajagopal:2015roa}, but also we have found the gravitational anomaly contribution to the chiral drag force which was neglected in computing \eqref{last_result} in \cite{Rajagopal:2015roa}.

 \section{Summary, Conclusion and Outlook}\label{8}
 In this paper we have studied a system of right- and left-handed Weyl  fermions in presence of a constant magnetic field, in the framework of chiral kinetic theory. Specifically we have computed the hydrodynamic excitations in this system. Not only the density of right- and left-handed charges may propagate, but also the energy and momentum perturbations are allowed to couple them and propagate in our system. To our knowledge, non of previous studies in the literature has taken the effect of momentum perturbations into account.

 As our first result we have shown that the true computations in chiral kinetic theory in such system requires a modification in the definition of the momentum current \eqref{momentum_us}. The latter is dictated by the Lorentz invariance. We have used this modified current to compute the linearized equations of momentum conservation in the hydro regime. Together with the linearized form of the energy and charge conservation equations, the latter make a set of six coupled equations \eqref{Mab}, \eqref{MatrixB1}. From these equations we have computed the set of hydro modes in the system \eqref{modes}.  We have seen that these modes are not the same as those computed directly from the hydrodynamic in the LL frame, before. This observation was our main motivation for the second part  of the paper.
	
	By comparing the energy flow in  the rest frame of our system with that in the rest frame of the fluid in the LL frame, we showed that 
	there is a boost linking between the rest frame of the fluid in the above two cases in equilibrium \eqref{general_boost}.  Let us mention that while somewhere in the  literature the change of the frame is referred to as the "shift in the velocity of the fluid" \cite{Landsteiner:2016led}, to our knowledge,  nowhere it was related to a boost between the rest frames of the fluid in two frames, specifically in equilibrium, before. 
	
	To confirm the idea of boost, we applied it to the set of dispersion relations  obtained from chiral kinetic theory and found the results of hydrodynamics in the LL frame  \eqref{boost_12}, \eqref{boost_34}, \eqref{boost_56}. We then  tested the idea in another direction. Inspired by the fact that the hydrodynamic frame in a system of single chirality Weyl fermions is a no-drag frame,  we found the structure of the hydro constitutive relations in  a no-drag frame for a system of non-Abelian fermions (see \eqref{energy-component}, \eqref{j-component} and \eqref{entropy-component}). The reason for choosing such system is two-fold. First, in a special case this system may reduce to the system of right- and left-handed  fermions discussed in the first part of the paper. Second, the constitutive relations for such system in the LL frame were computed before \cite{Neiman:2010zi}. Focusing on the results of the no-drag frame, we then found  the velocity of boost which transforms the rest frame of the fluid in no-drag frame to that in LL frame \eqref{boost_non_Abelian}.
Using this we could reproduce the results previously found in the LL frame \eqref{trans-Lan}.

Our last result was to reproduce one of the well-known results of AdS/CFT, however, without performing any holographic computations. 
In \cite{Rajagopal:2015roa}, the chiral drag force exerted on a  quark moving in a chiral fluid was computed. We first determined the most general form of the chiral transport in a chiral fluid in presence of constant magnetic field and vorticity up to first order in derivative expansion  \eqref{general_F}. Then we evaluated it for a quark in the rest frame of the fluid in LL frame. By boosting the force from LL to no-drag frame and demanding the resultant force to vanish in the rest frame of the fluid for a quark at rest, we found two constraints on the coefficients of general drag force \eqref{constraint_1}, \eqref{constraint_2}. Using this we could reproduce the chiral drag force exerted on a quark at rest in the rest frame of the fluid in LL frame \eqref{last_result}.
	
Let us now discuss on the physical implications of the results.
Concerning the first part of the paper, and specifically the modification of the momentum current,  we think that it might be an important step towards understanding the collisions in chiral kinetic theory in presence of magnetic field. The lack of such framework was pointed out in  \cite{Chen:2015gta}.
	It can be also important when considering the chiral kinetic theory beyond the hydrodynamic limit. The latter means that such modification is crucial to find the precise spectrum of the non-hydrodynamic modes in the system. 
	
	 The second part of the paper is mostly involved with the notion of frame in hydrodynamics. 
	 What we have shown in this paper may have some important consequences. First let us recall that the chiral Alfv\'en wave was found firstly as a  new gapless excitation in a fluid with one single chirality, in the LL frame \cite{Yamamoto:2015ria,Abbasi:2015saa,Kalaydzhyan:2016dyr}.  Interestingly if we repeat the computations in the Lab frame we will see that at zero density limit, the chiral Alfv\'en wave basically does not exist in this frame. The reason is that the velocity of boost between two frames is exactly the velocity of this wave in LL frame (see Appendix \ref{App_3}). This means that the appearance of chiral Alfv\'en wave in the system of single chirality fermions, in LL frame, is a frame-dragging effect. However, it is worth-mentioning that such wave would propagate in the Lab frame if we had two species of fermions with different densities \cite{Abbasi:2016rds}. The latter may physically be related to the QCD fluid in heavy ion collision experiments.
	 
	 Second, our results show that in general,  the hydrodynamic modes are not frame invariant, while they seem physically to be so. They are in fact frame invariant as long as no non-vanishing one derivative vector exists in equilibrium. It is exactly for this reason that in a non-chiral fluid, the set of hydrodynamic modes in LL frames are the same as those in Eckart frame. It has to be noted the latter would no longer be true in a strong magnetic field; since in this case the magnetic field can turn on the off-diagonal components of the energy momentum tensor and  contribute to the boost velocity, even in a non-chiral matter \cite{Jensen:2011xb,Ammon:2017ded}. In summary what precisely determines whether the hydro modes  are frame dependent or not is the presence or non-presence of a non-vanishing vector in  equilibrium in the system.

	\section{Acknowledgement}\label{Acknow}
	We would like to thanks Ali Davody for valuable discussions and collaboration in the early stages of this work. We would also like to thank  D. Allahbakhshi, M. Chernodub and S.F. Taghavi for discussion. N. A. would like to thank M. Mohammadi for the support of the theory team. K. N. would like to thank Professor H. Arafei for encouragements and supports. K. N. would also like to thank school of Particles and Accelerators of IPM.
\section{Appendix}\label{suseptibility}
\subsection{Thermodynamic relations}
		To compare obtained modes from \eqref{modes} with those of  hydrodynamics approach, it is useful to introduce some thermodynamic quantities. In the case of double- current fluid, 
	for small deviations from  equilibrium state, one can write the changes of macroscopic quantities in terms of the fluctuations of some independent variables 
	\begin{eqnarray}
	&&\delta\epsilon= \left(\frac{\partial \epsilon}{\partial T}\right)_{\mu_{{R}}, \mu_{{L}}} \delta T + \left(\frac{\partial \epsilon}{\partial \mu_{R}}\right)_{T, \mu_{{L}}} \delta \mu_{R} +\left(\frac{\partial \epsilon}{\partial \mu_{L}}\right)_{\mu_{{R}}, T} \delta \mu_{L},\nonumber\\
	&& \delta n_{R}= \left(\frac{\partial n_{R}}{\partial T}\right)_{\mu_{{R}}, \mu_{{L}}} \delta T + \left(\frac{\partial n_{R}}{\partial \mu_{R}}\right)_{T, \mu_{{L}}} \delta \mu_{R} +\left(\frac{\partial n_{R}}{\partial \mu_{L}}\right)_{\mu_{{R}}, T} \delta \mu_{L},\nonumber\\
	&& \delta n_{L}= \left(\frac{\partial n_{L}}{\partial T}\right)_{\mu_{{R}}, \mu_{{L}}} \delta T + \left(\frac{\partial n_{L}}{\partial \mu_{R}}\right)_{T, \mu_{{L}}} \delta \mu_{R} +\left(\frac{\partial n_{L}}{\partial \mu_{L}}\right)_{\mu_{{R}}, T} \delta \mu_{L}.
	\end{eqnarray}
	By using the given expressions in equation (\ref{thermo}), one could calculate the above coefficients. However,  we define these coefficients (susceptibility matrix elements) in the basis of vector- axial  as follow
	\begin{equation}\label{}
	\varepsilon=\begin{pmatrix}
	\alpha_{1}&& \alpha_{2}&& \alpha_{3}\\
	\beta_{1}&& \beta_{2}&& \beta_{3}\\
	\gamma_{1}&& \gamma_{2}&& \gamma_{3}
	\end{pmatrix}=
	\begin{pmatrix}
	\frac{\partial \epsilon}{\partial T}&&  \frac{\partial \epsilon}{\partial \mu_{V}}&&  \frac{\partial \epsilon}{\partial \mu_{A}}\\
	\frac{\partial n_{V}}{\partial T}&&  \frac{\partial n_{V}}{\partial \mu_{V}}&&  \frac{\partial n_{V}}{\partial \mu_{A}}\\
	\frac{\partial n_{A}}{\partial T}&&  \frac{\partial n_{A}}{\partial \mu_{V}}&& \frac{\partial n_{A}}{\partial \mu_{A}}
	\end{pmatrix}.
	\end{equation}

	\subsection{Boost transformation from No-Drag frame to LL Frame}
	\label{ppendix_boost}
For the sake of simplicity we get the boost in the third direction only. The energy momentum components are transformed as the following
	\begin{eqnarray}\nonumber
	[T]^{\mu\nu}&=&[\Lambda_{\boldsymbol{\beta}}]^{\mu}_{\,\,\,\alpha}\,[T']^{\alpha\gamma}\,[\Lambda_{\boldsymbol{\beta}}]_{\gamma}^{\,\,\,\,\nu}\\\nonumber
	\begin{pmatrix}
	T^{00}  & T^{0x}& T^{0y}&T^{0z} \\
	T^{x0} &T^{xx}&T^{xy}&T^{xz}\\
	T^{y0}  & T^{yx}& T^{yy}&T^{yz} \\
	T^{z0} &T^{zx}&T^{zy}&T^{zz}\\\end{pmatrix}&=&\begin{pmatrix}
	\gamma  & 0& 0&-\gamma \beta\,\epsilon_f \\
	0 & 1&0&0\\
	0&0&1&0\\
	-\gamma \beta\,\epsilon_f&0&0&\gamma\\
	\end{pmatrix} \begin{pmatrix}
	T'^{00} & T'^{0x} & T'^{0y} &T'^{0z} \\
	T'^{x0} & T'^{xx} & T'^{xy} &T'^{xz}\\
	T'^{y0} & T'^{yx} & T'^{yy} &T'^{yz} \\
	T'^{z0} &T'^{zx}  & T'^{zy} &T'^{zz}\\\end{pmatrix} \begin{pmatrix}
	\gamma  & 0& 0&-\gamma \beta\,\epsilon_f \\
	0 & 1&0&0\\
	0&0&1&0\\
	-\gamma \beta\,\epsilon_f&0&0&\gamma
	\end{pmatrix}\nonumber\\
	&=&\begin{pmatrix}
	\gamma  & 0& 0&-\gamma \beta\,\epsilon_f \\
	0 & 1&0&0\\
	0&0&1&0\\
	-\gamma \beta\,\epsilon_f&0&0&\gamma\\
	\end{pmatrix}
	\begin{pmatrix}
	\epsilon  & 0& 0&w \beta \epsilon_f\\
	0 &p&0&0\\
	0  & 0& p&0 \\
	w \beta \epsilon_f&0&0&p\\\end{pmatrix}
	\begin{pmatrix}
	\gamma  & 0& 0&-\gamma \beta\,\epsilon_f \\
	0 & 1&0&0\\
	0&0&1&0\\
	-\gamma \beta\,\epsilon_f &0&0&\gamma
	\end{pmatrix}\nonumber\\
	&=& \begin{pmatrix}
	\epsilon  & 0& 0&0 \\
	0 &p&0&0\\
	0  & 0& p&0 \\
	0 &0&0&p\\\end{pmatrix}+O(\epsilon_{f}^2).
	\end{eqnarray}
	Here $w$ is enthalpy $w=\epsilon + p$ and $\beta$ is the velocity of boost $\beta= \frac{\xi_{TB}^a B^a}{w} + \frac{\xi_{T\omega} \Omega}{w}$.  As  expected we find no energy flows in the rest frame of the LL frame
	\begin{equation}\label{Landau-components}
	T^{00}=\epsilon,\,\,\,\,\,\,T^{0i}=0,\,\,\,\,\,T^{ij}=p\delta^{ij}.
	\end{equation}
	In what follows, we do this transformation for currents
	\begin{eqnarray}\nonumber
	[J^{a}]^{\mu}&=& [\Lambda_{\boldsymbol{\beta}}]^{\mu}_{\,\,\,\alpha} \,[J'^{a}]^{\alpha}\\
	\begin{pmatrix}
	(J^0)^a \\
	(J^{x})^a \\
	(J^{y})^a\\
	(J^{z})^a\\
	\end{pmatrix}	&=&\begin{pmatrix}
	\gamma  & 0& 0&-\gamma \beta\,\epsilon_f \\
	0 & 1&0&0\\
	0&0&1&0\\
	-\gamma \beta\,\epsilon_f&0&0&\gamma\\
	\end{pmatrix}\begin{pmatrix}
	(J'^0)^a \\
	(J'^{x})^a \\
	(J'^{y})^a\\
	(J'^{z})^a\\
	\end{pmatrix}=\,\begin{pmatrix}
	\gamma  & 0& 0&-\gamma \beta\,\epsilon_f \\
	0  & 1&0&0\\
	0&0&1&0\\
	-\gamma \beta\,\epsilon_f&0&0&\gamma\\
	\end{pmatrix}\begin{pmatrix}
	n^{a} \\
	0\\
	0\\
	\xi_{JB}^{ab} B^{b} + \xi_{J\omega}^a \Omega\\
	\end{pmatrix}\nonumber\\
	&=&\begin{pmatrix}
	n^{a} \\
	0\\
	0\\
	(\xi_{JB}^{ab} -\frac{ \xi_{TB}^b n^a}{w}) B^{b} + (\xi_{J\omega}^a - \frac{\xi_{T\omega} n^a}{w}) \Omega\\
	\end{pmatrix}+O(\epsilon_f)^2.
	\end{eqnarray}
 We do the same calculation for entropy current
	\begin{eqnarray}\nonumber
	[S]^{\mu}&=& [\Lambda_{\boldsymbol{\beta}}]^{\mu}_{\,\,\,\alpha} \,[S']^{\alpha}\\
	\begin{pmatrix}
	S^0 \\
	S^{x} \\
	S^{y}\\
	S^{z}\\
	\end{pmatrix}	&=&\begin{pmatrix}
	\gamma  & 0& 0&-\gamma \beta\,\epsilon_f \\
	0 & 1&0&0\\
	0&0&1&0\\
	-\gamma \beta\,\epsilon_f&0&0&\gamma\\
	\end{pmatrix}\begin{pmatrix}
	S'^0 \\
	S'^{x} \\
	S'^{y}\\
	S'^{z}\\
	\end{pmatrix}=\begin{pmatrix}
	\gamma  & 0& 0&-\gamma \beta\,\epsilon_f \\
	0  & 1&0&0\\
	0&0&1&0\\
	-\gamma \beta\,\epsilon_f&0&0&\gamma\\
	\end{pmatrix}\begin{pmatrix}
	s \\
	0\\
	0\\
	\xi_{sB}^{a} B^{a} + \xi_{s\omega} \Omega\\
	\end{pmatrix}\nonumber\\\nonumber
	\\
		&=&\begin{pmatrix}
	s \\
	0\\
	0\\
	(\xi_{sB}^{a} -\frac{s\, \xi_{TB}^a }{w}) B^{a} + (\xi_{s\omega} - \frac{s\,\xi_{T\omega}}{w}) \Omega\\
	\end{pmatrix}+O(\epsilon_f)^2.
	\end{eqnarray}
	By these components one could write the general form of entropy current in LL frame as
	\begin{eqnarray}\label{SLan1}
		S^{\mu}_{LL}= s u^{\mu} + (\xi_{sB}^{a} -\frac{s\, \xi_{TB}^a }{w}) B^{a \mu} + (\xi_{s\omega} - \frac{s\,\xi_{T\omega} }{w}) \omega^{\mu}.
	\end{eqnarray}
	In LL frame the entropy current has the following form  \cite{Son:2009tf},\cite{Neiman:2010zi}
	\begin{eqnarray}\label{SLan2}
		S^{\mu}_{LL} = s u^{\mu} - \frac{\mu^a}{T} J^{a \mu}_{Lan} + \tilde{\xi}^{a}_B B^{a\mu} + \tilde{\xi}_{\omega} \omega^{\mu}.
	\end{eqnarray}
Equating  relations \eqref{SLan1} and \eqref{SLan2}
	 and  by using   relations \eqref{energy-component}, \eqref{j-component}, \eqref{entropy-component} and \eqref{trans-Lan} 
	we obtain the transport coefficients $\tilde{\xi}^a_B$ and $\tilde{\xi}_\omega$ as
	\begin{eqnarray}
		 \tilde{\xi}^{a}_B &=&\xi_{sB}^{a} -\frac{ s\,\xi_{TB}^a}{w} + \xi_{B\, Lan}^{ab} \frac{\mu^b}{T}=\frac{1}{2 T} C^{abc} \mu^b \mu^c + \frac{\beta^a}{2} T,\nonumber\\
		\tilde{\xi}_\omega&=&\xi_{s\omega} -\frac{ s\,\xi_{T\omega} }{w} + \xi_{\omega\, Lan}^{a} \frac{\mu^a}{T}=\frac{1}{3 T} C^{abc} \mu^a \mu^b \mu^c +  \beta^{a} \mu^{a} T + \frac{\gamma}{3} T^{2},
	\end{eqnarray} 
	which are exactly those obtained in \cite{Neiman:2010zi}.
	\begin{equation}
	\begin{split}
	J'^{0}=&\,J^{0}+\,\beta\, J^{x}\,\epsilon_f+O(\epsilon_f^2)
	\\
	=&(n\,u^0+\xi_B B^0\,\epsilon_f)+\,\frac{\bar{\boldsymbol{\chi}}_{R}-\bar{\boldsymbol{\chi}}_{L}}{4w}\text{B}^x\,(n\,u^x)\epsilon_f+O(\epsilon_f^2)\\
	=&(n\,u^0+\xi_B B^0\,\epsilon_f)+\,n\,\frac{\bar{\boldsymbol{\chi}}_{R}-\bar{\boldsymbol{\chi}}_{L}}{4 w}\,B^0\epsilon_f+O(\epsilon_f^2)\\
	=&n\,u^0+\,\left(\xi_B+\,n\,\frac{\bar{\boldsymbol{\chi}}_{R}-\bar{\boldsymbol{\chi}}_{L}}{4 w}\right)\,B^0\epsilon_f+O(\epsilon_f^2)
	\end{split}
	\end{equation}
	\begin{equation}
	\begin{split}
	J'^{x}=&\,J^{x}+\,\beta\, J^{0}\,\epsilon_f+O(\epsilon_f^2)
	\\
	=&(n\,u^x+\xi_B B^x\,\epsilon_f)+\,\frac{\bar{\boldsymbol{\chi}}_{R}-\bar{\boldsymbol{\chi}}_{L}}{4w}\text{B}^x\,(n\,u^0)\epsilon_f+O(\epsilon_f^2)\\
	=&(n\,u^x+\xi_B B^x\,\epsilon_f)+\,n\,\frac{\bar{\boldsymbol{\chi}}_{R}-\bar{\boldsymbol{\chi}}_{L}}{4 w}\,B^x\epsilon_f+O(\epsilon_f^2)\\
	=&n\,u^x+\,\left(\xi_B+\,n\,\frac{\bar{\boldsymbol{\chi}}_{R}-\bar{\boldsymbol{\chi}}_{L}}{4 w}\right)\,B^x\epsilon_f+O(\epsilon_f^2)
	\end{split}
	\end{equation}
	Let us review the above points. We showed that in a weak magnetic field $\textbf{B}$ and in the absence of dissipation, the LL frame is a Lorentz frame which is moving with velocity \eqref{v_boost} with respect to the Laboratory frame. 
	
	\subsection{Comparison Between Hydro Modes in LL and Lab Frame for a Fluid with a Single  Chirality}
	\label{App_3}
		For a conformal fluid in  the laboratory frame:
			\begin{eqnarray}
			\label{Tmunu}
			T^{\mu \nu} &= & w u^{\mu} u^{\nu} + p g^{\mu \nu}+\sigma^{\mathcal{B}}_{\epsilon}(u^{\mu}B^{\nu}+u^{\nu}B^{\mu})+\sigma^{\mathcal{V}}_{\epsilon}(u^{\mu}\omega^{\nu}+u^{\nu}\omega^{\mu}),\\
			\label{jmu}
			j^{\mu} & = & nu^{\mu} + \sigma^{\mathcal{V}} \omega^{\mu}+\sigma^{\mathcal{B}} B^{\mu},
			\end{eqnarray}
			the non-dissipative transport coefficients are then given as the following \cite{Landsteiner:2012kd}
			\begin{eqnarray}
			\sigma^{\mathcal{B}}&=&\frac{\mu}{4 \pi^2}\,\,\,\,\,\,\,\,\sigma^{\mathcal{V}}=\frac{\mu^2}{4 \pi^2}+\frac{T^2}{12}\\
			\sigma^{\mathcal{B}}_{\epsilon}&=&\frac{1}{2}\,\sigma^{\mathcal{V}}\,\,\,\,\,\,\,\sigma^{\mathcal{V}}_{\epsilon}=\frac{\mu^3}{6 \pi^2}+\frac{\mu T^2}{6}.
			\end{eqnarray}
			In the LL frame we have
			\begin{eqnarray}
			T_{LL}^{\mu\nu}&=&(\epsilon+p)u^{\mu}u^{\nu}+p \eta^{\mu\nu}\\
			J_{LL}^{\mu}&=&n u^{\mu}+ \underbrace{\left(\sigma^{\mathcal{B}}-\frac{n}{\epsilon+p}\sigma^{\mathcal{B}}_{\epsilon}\right)}_{\xi_{B}}B^{\mu}+ \underbrace{\left(\sigma^{\mathcal{V}}-\frac{n}{\epsilon+p}\sigma^{\mathcal{V}}_{\epsilon}\right)}_{\xi}\omega^{\mu}
			\end{eqnarray}
	It is not hard to see that the set of hydro modes in the above two frames may be given as in the Table \ref{table_2}
			\begin{table}[!htb]
				\begin{center}
					\begin{tabular}{|c|c|c|}
						\hline
						\hline
						Type of mode & Lab frame & Landau-lifshitz frame \\
						\hline
						\hline
						&  & \\ 
						
						CMHW    & $v_1=\left(\frac{-\alpha_1\big(\frac{\partial \sigma^{\mathcal{B}}}{\partial \mu}-\frac{n}{w}\frac{\partial \sigma^{\mathcal{B}}_{\epsilon}}{\partial\mu}\big)+\alpha_2\big(\frac{\partial \sigma^{\mathcal{B}}}{\partial T}-\frac{n}{w}\frac{\partial \sigma^{\mathcal{B}}_{\epsilon}}{\partial T}\big)}{{\alpha_2 \beta_1-\alpha_1 \beta_2}}\right) B$&  $v_1=\left(\frac{-\alpha_1 \frac{\partial \xi_B}{\partial \mu}+\alpha_2\frac{\partial \xi_B}{\partial T}}{\alpha_2 \beta_1-\alpha_1 \beta_2}\right) B $   \\
						
						&   &  \\
						\hline
						&  & \\ 
						
						Sound     & $v_{2,3}=\pm c_s-\frac{c_s^2-1}{ w}\sigma^{\mathcal{B}}_{\epsilon} B$ &  $v_{2,3}=\pm c_s$  \\
						
						&   &  \\
						\hline
						&  & \\ 
						
						CAW     & $v_{4,5}=\frac{n}{2w^2}\sigma^{\mathcal{V}}_{\epsilon} B$&  $v_{4,5}=-\frac{\xi }{2 w}$B  \\
						
						&   &  \\
						\hline
						\hline
					\end{tabular}
				\end{center}
				\caption{Hydro modes in Lab frame versus those in LL frame}
				\label{table_2}
			\end{table}
		One can see that the above two sets of hydro modes are related to each other with the following boost velocity
			\begin{equation}
 \beta^i=-\frac{1}{w}(\sigma_{\epsilon}^{\mathcal{B}}B^i+\frac{}{}\sigma_{\epsilon}^{\mathcal{V}}\Omega^{i})
		\end{equation}
		Interestingly, when $\mu=0$, we obtain
		\begin{equation}
		\boldsymbol{v}_{4,5}^{Lab}=0,\,\,\,\,\,\,\,\,\,\,\,\,\,\,\,\boldsymbol{v}_{4,5}^{LL}=\boldsymbol{\beta}.
		\end{equation}
		\subsection{Conservation Equations Integrated Over Unit Sphere}
		\label{App_4}
		In the relations \eqref{coeff_jR}, \eqref{coeff-energy}, \eqref{coeff-mompar} and \eqref{coeff-momperp} we gave the set of expressions contributing to the integrands in the condservation equations.  In order to derive the corresponding hydro equations  \eqref{current-right} and \eqref{current-left}, \eqref{en},  \eqref{mom-parallel}, \eqref{mom-perp1} and \eqref{mom-perp2}, we have to perform momentum integrals and then sum over particle and anti-particle contributions. The computations are cumbersome, however,  in this Appendix we give the expressions after peforming the integral over the unit sphere in momentum space. In all expressions we use
		\begin{equation}
			\mathbf{B} \cdot \mathbf{p} = \text{B} \, \text{p}\,\cos\theta,\,\,\,\,\,\,\,\,\,\,\mathbf{p} \cdot \boldsymbol{k} = \text{p} \, k\,\cos\theta,\,\,\,\,\,\,\,\,\,\,\,\,\mathbf{B} \cdot \boldsymbol{k} = \text{B} \, k.
		\end{equation}
	Let us start by performing the integral in the expressions of \eqref{coeff_jR}. One obtains
		  \begin{eqnarray}
		  	&&\int \frac{d\Omega}{(2\pi)^3}\, e\, \mathcal{A}_{\delta \mu_{{\chi'}}}^{C_{\chi}}= -\,\frac{i\,\beta\,e^{\beta(\text{p} - e \mu_{\chi})}} {6\pi^2 (e^{\beta(\text{p} - e \mu_{\chi})}+1)^2} \left\{\,3\text{p}^2 \omega- \lambda \text{B}\, k\, \left(2 + \text{p}\, \beta\, \tanh(\frac{\beta}{2}(\text{p} - e \mu_{\chi}))\right)\right\},\nonumber\\
		  	&&\int \frac{d\Omega}{(2\pi)^3}\, e\, \mathcal{A}_{\delta \beta}^{C_{\chi}}= \frac{i\,e\, (\text{p} - e\mu_{\chi})\,e^{\beta(\text{p} - e \mu_{\chi})}} {6\pi^2 (e^{\beta(\text{p} - e \mu_{\chi})}+1)^2} \left\{\,3\text{p}^2 \omega- \lambda \text{B}\, k\, \left(\frac{\text{p} - 2e\mu_{\chi}}{\text{p} - e\mu_\chi} + \text{p}\, \beta\, \tanh(\frac{\beta}{2}(\text{p} - e \mu_{\chi}))\right)\right\},\nonumber\\
		  	&&\int \frac{d\Omega}{(2\pi)^3}\, e\, \mathcal{A}_{\pi_{{i}}^{\perp}}^{C_{\chi}}=0,\nonumber\\
		  	&&\int \frac{d\Omega}{(2\pi)^3}\, e\, \mathcal{A}_{\pi^{\parallel}}^{C_{\chi}}= \frac{i\,\beta\,e\,\text{p}\,e^{\beta(\text{p} - e \mu_{\chi})}} {6\pi^2 w\,(e^{\beta(\text{p} - e \mu_{\chi})}+1)^2} \left\{k\,\text{p}^2 - \lambda \text{B}\, \omega\, \left(1 + \text{p}\, \beta\, \tanh(\frac{\beta}{2}(\text{p} - e \mu_{\chi}))\right)\right\}.
		  \end{eqnarray}
		 For the expressions in \eqref{coeff-energy} we reach
		  \begin{eqnarray}
		  	&&\int \frac{d\Omega}{(2\pi)^3}\, \mathcal{A}_{\delta \mu_{{\chi}}}^{\text{E}}= -\,\frac{i\,\beta\,e\,\text{p}\,e^{\beta(\text{p} - e \mu_{\chi})}} {6\pi^2 (e^{\beta(\text{p} - e \mu_{\chi})}+1)^2} \left\{\,3\text{p}^2 \omega- \lambda \text{B}\, k\, \left(1 + \text{p}\, \beta\, \tanh(\frac{\beta}{2}(\text{p} - e \mu_{\chi}))\right)\right\},\nonumber\\
		  	&&\int \frac{d\Omega}{(2\pi)^3}\, \mathcal{A}_{\delta \beta}^{\text{E}}= \frac{i\,\text{p}\, (\text{p} - e\mu_{\chi})\,e^{\beta(\text{p} - e \mu_{\chi})}} {6\pi^2 (e^{\beta(\text{p} - e \mu_{\chi})}+1)^2} \left\{\,3\text{p}^2 \omega- \lambda \text{B}\, k\, \left(-\,\frac{e\mu_{\chi}}{\text{p} - e\mu_\chi} + \text{p}\, \beta\, \tanh(\frac{\beta}{2}(\text{p} - e \mu_{\chi}))\right)\right\},\nonumber\\
		  	&&\int \frac{d\Omega}{(2\pi)^3}\, \mathcal{A}_{\pi_{{i}}^{\perp}}^{\text{E}}=0,\nonumber\\
		  	&&\int \frac{d\Omega}{(2\pi)^3}\, \mathcal{A}_{\pi^{\parallel}}^{\text{E}}= \frac{i\,\beta\,\text{p}^3\,e^{\beta(\text{p} - e \mu_{\chi})}} {6\pi^2 w\,(e^{\beta(\text{p} - e \mu_{\chi})}+1)^2} \left\{k\,\text{p} - \lambda \text{B}\, \omega\,   \beta\, \tanh(\frac{\beta}{2}(\text{p} - e \mu_{\chi}))\right\}.
		  \end{eqnarray}
		 Analogous computation for \eqref{coeff-mompar} gives
		  \begin{eqnarray}
		  	&&\int \frac{d\Omega}{(2\pi)^3}\, \mathcal{A}_{\delta \mu_{{\chi}}}^{\text{P}^\parallel}= \frac{i\,\beta\,e\, \text{p}\, e^{\beta(\text{p} - e \mu_{\chi})}} {6\pi^2 (e^{\beta(\text{p} - e \mu_{\chi})}+1)^2} \left\{k\,\text{p}^2 - \lambda \text{B}\, \omega\, \left(1 + \text{p}\, \beta\, \tanh(\frac{\beta}{2}(\text{p} - e \mu_{\chi}))\right)\right\},\nonumber\\
		  	&&\int \frac{d\Omega}{(2\pi)^3}\, \mathcal{A}_{\delta \beta}^{\text{P}^\parallel}= -\frac{i\,\text{p}\, (\text{p} - e\mu_{\chi})\,e^{\beta(\text{p} - e \mu_{\chi})}} {6\pi^2 (e^{\beta(\text{p} - e \mu_{\chi})}+1)^2} \left\{k\,\text{p}^2 - \lambda \text{B}\, \omega\, \left(-\,\frac{e\mu_{\chi}}{\text{p} - e\mu_\chi} + \text{p}\, \beta\, \tanh(\frac{\beta}{2}(\text{p} - e \mu_{\chi}))\right)\right\},\nonumber\\
		  	&&\int \frac{d\Omega}{(2\pi)^3}\, \mathcal{A}_{\pi_{{i}}^{\perp}}^{\text{P}^\parallel}=0,\nonumber\\
		  	&&\int \frac{d\Omega}{(2\pi)^3}\, \mathcal{A}_{\pi^{\parallel}}^{\text{P}^\parallel}=-\, \frac{i\,\beta\,\text{p}^2\,e^{\beta(\text{p} - e \mu_{\chi})}} {30\pi^2 w\,(e^{\beta(\text{p} - e \mu_{\chi})}+1)^2} \left\{5\,\text{p}^2 \omega - 3\lambda \text{B}\, k\, \left(2 + \text{p}\, \beta\, \tanh(\frac{\beta}{2}(\text{p} - e \mu_{\chi}))\right)\right\}.
		  \end{eqnarray}
		  For the momentum conservation equation \eqref{coeff-momperp},  we may write the expressions corresponded to two transverse directions separately. For $\text{P}^{\perp}_{1}$ we obtain
		  \begin{eqnarray}
		  	&&\int \frac{d\Omega}{(2\pi)^3} \mathcal{A}_{\delta \mu_{{\chi}}}^{\text{P}_1^{\perp}}=0,~~\int \frac{d\Omega}{8\pi^3} \mathcal{A}_{\delta \beta}^{\text{P}_1^{\perp}}=0,~~\int \frac{d\Omega}{8\pi^3} \mathcal{A}_{\delta \pi^\parallel}^{\text{P}_1^{\perp}}=0,\nonumber\\
		  	&&\int \frac{d\Omega}{(2\pi)^3}\, \mathcal{A}_{\pi_{{1}}^{\perp}}^{\text{P}_1^\perp}=-\, \frac{i\,\beta\,\text{p}^2\,e^{\beta(\text{p} - e \mu_{\chi})}} {30\pi^2 w\,(e^{\beta(\text{p} - e \mu_{\chi})}+1)^2} \left\{5\,\text{p}^2 \omega +\lambda \text{B}\, k\, \left(3 - \text{p}\, \beta\, \tanh(\frac{\beta}{2}(\text{p} - e \mu_{\chi}))\right)\right\},\nonumber\\
		  	&&\int \frac{d\Omega}{(2\pi)^3}\, \mathcal{A}_{\pi_{{2}}^{\perp}}^{\text{P}_1^\perp}=-\, \frac{\beta\,e\,\text{p}^3\,(\text{B} + \lambda \,k\, \omega)e^{\beta(\text{p} - e \mu_{\chi})}} {6\pi^2 w\,(e^{\beta(\text{p} - e \mu_{\chi})}+1)^2},
		  \end{eqnarray}  
		  and for $\text{P}_2^{\perp}$ 
		  \begin{eqnarray}
		  &&\int \frac{d\Omega}{(2\pi)^3} \mathcal{A}_{\delta \mu_{{\chi}}}^{\text{P}_2^{\perp}}=0,~~\int \frac{d\Omega}{8\pi^3} \mathcal{A}_{\delta \beta}^{\text{P}_2^{\perp}}=0,~~\int \frac{d\Omega}{8\pi^3} \mathcal{A}_{\delta \pi^\parallel}^{\text{P}_2^{\perp}}=0,\nonumber\\
		  &&\int \frac{d\Omega}{(2\pi)^3}\, \mathcal{A}_{\pi_{{1}}^{\perp}}^{\text{P}_2^\perp}=\frac{\beta\,e\,\text{p}^3\,(\text{B} + \lambda \,k\, \omega)e^{\beta(\text{p} - e \mu_{\chi})}} {6\pi^2 w\,(e^{\beta(\text{p} - e \mu_{\chi})}+1)^2},\nonumber\\
		  &&\int \frac{d\Omega}{(2\pi)^3}\, \mathcal{A}_{\pi_{{2}}^{\perp}}^{\text{P}_2^\perp}=-\, \frac{i\,\beta\,\text{p}^2\,e^{\beta(\text{p} - e \mu_{\chi})}} {30\pi^2 w\,(e^{\beta(\text{p} - e \mu_{\chi})}+1)^2} \left\{5\,\text{p}^2 \omega +\lambda \text{B}\, k\, \left(3 - \text{p}\, \beta\, \tanh(\frac{\beta}{2}(\text{p} - e \mu_{\chi}))\right)\right\}.
		  \end{eqnarray} 

	\bibliographystyle{utphys}
	
	\providecommand{\href}[2]{#2}\begingroup\raggedright\endgroup

\end{document}